\documentclass[twocolumn, aps, pra,]{revtex4-2}

\usepackage[colorlinks=true, citecolor=blue, urlcolor=blue, linkcolor=magenta]{hyperref}
\usepackage{braket}
\usepackage{amsmath,amssymb,amsthm}
\usepackage{easybmat}
\usepackage[colorlinks=true,citecolor=blue,urlcolor=blue, linkcolor=magenta]{hyperref}
\usepackage[pdftex]{graphicx}
\usepackage{times,txfonts}
\usepackage{braket}
\usepackage{color}
\usepackage{natbib}
\usepackage{appendix}

\begin{document}
    \title{Strong coupling non-Markovian quantum thermodynamics of a finite-bath system}
\author{Devvrat Tiwari\textsuperscript{}}
\email{devvrat.1@iitj.ac.in}
\author{Baibhab Bose\textsuperscript{}}
\email{baibhab.1@iitj.ac.in}
\author{Subhashish Banerjee\textsuperscript{}}
\email{subhashish@iitj.ac.in}
\affiliation{Indian Institute of Technology Jodhpur-342030, India\textsuperscript{}}

\date{\today}

\begin{abstract}
    The focus is on understanding the quantum thermodynamics of strongly coupled non-Markovian quantum systems. To this end, a non-trivial, non-Markovian model of a central spin surrounded by a spin bath is taken up, and its exact evolution is derived for arbitrary system-bath couplings. The fundamental quantum thermodynamic quantities, such as system and bath internal energies, work, heat, entropy production, and ergotropy, are calculated using the dynamics and original system (bath) Hamiltonian. An explicit expression for the work, a mismatch between the system and bath internal energies, is derived. The thermodynamic entropy of the system at thermal equilibrium is studied using the Hamiltonian of mean force in the strong coupling regime. The role of a canonical Hamiltonian in calculating the above thermodynamic quantities, a recently developed technique, is also investigated. Further, an interesting observation relevant to the spin bath acting as a charger is made in a scenario where the central spin is envisaged as a quantum battery.
\end{abstract}

\maketitle
\section{Introduction}
\label{intro}
A reliable description of quantum systems involves the interaction with their surroundings. These interactions substantially impact the dynamics of the system. The theory of open quantum systems~\cite{7t26_6_weiss2012quantum, BreurPoqs, banerjee2018open} provides a framework for studying such systems. Quantum systems that can perform quantum information 
theoretic tasks, for example, trapped ions~\cite{trapped_ions}, quantum dots~\cite{Qdots}, NMR qubits~\cite{nmr_qubits}, Josephson junctions~\cite{Josesph}, and many more, are subjected to environmental interactions. The theory of open quantum systems has found numerous applications in quantum information and its interface with other aspects of quantum physics~\cite{7t26_1,7t26_2,7t26_3_lindblad1976generators,7t26_4_CALDEIRA1983374,7t26_5_FEYNMAN1963118,7t26_6_weiss2012quantum,7t26_7_10.1143/PTP.20.948,7t26_8_10.1063/1.1731409,7t26_9_PhysRevA.62.042105,7t26_10_PhysRevE.67.056120,7t26_11_PhysRevA.78.052316,7t26_13_PhysRevE.97.062108,7t26_14_PhysRevD.99.095001,7t26_16_PhysRevD.97.053008,7t26_17_dixit2019study, Omkar2016}.
 
With impressive advancements in technology, rapid inroads have been made into the development of quantum technologies and quantum devices~\cite{Gemmer2009, Binder_book, sai_janet_book, KUMAR2023128832}. The impact of open system effects on these is indelible. The principles of quantum thermodynamics play an important role in these developments~\cite{kosloff_2022,landi_RevModPhys.93.035008, deffner_book}. The theoretical emphasis is to place the foundations of thermodynamics on a firm footing in the quantum regime~\cite{Alicki_1979, kosloff_2022, Hanggi_RevModPhys.92.041002, Lahiri2021}. On the technological and experimental front, these ideas can be used in the understanding and implementation of devices such as quantum batteries and heat engines~\cite{Alicki_battery, Binder_book, Binder_2015, VanHorne2020, campaioli, impact_devvrat_SB, bhanja_devvrat_SB, Nori_heat_engine, Allahverdyan_Ramandeep, TSMahesh_2022, TSMahesh_2019, Ramandeep_2009, Shubhrangshu_2021}. In this context, traditional involvement of open system ideas has been made using the Gorini-Kossakowski-Lindblad-Sudarshan (GKLS) dynamical evolution~\cite{7t26_2, 7t26_3_lindblad1976generators}. However, the development of theory, as well as experiment, demands pushing into regimes beyond that governed by the GKLS evolution, {\it viz.}, where strong-coupling non-Markovian (NM) effects play a significant role. 

Non-Markovian open quantum systems exemplify the delicate relationship between a quantum system and its surroundings. These systems, unlike their Markovian counterparts, retain a memory of their past, leading to a rich dynamical behavior~\cite{vega_alonso, 24t50_48_Utagi2020, 24t50_50_PhysRevLett.103.210401, 24t50_29_PhysRevA.84.052118, 24t50_27_PhysRevLett.105.050403, 24t50_47_doi:10.1142/S1230161218500142}. Many of these scenarios do not have a clear distinction between system and environment timelines. A prominent signature of non-Markovian evolution is the revival of quantum properties~\cite{24t50_50_PhysRevLett.103.210401}, the analysis of which is vital for a fundamental understanding of the system's dynamics at arbitrary coupling with the environment. The study of the dynamics of thermodynamic quantities in the presence of strong system-bath coupling in the non-Markovian regime is a challenging task~\cite{Strasberg_2016, Zhang_2022, 7t26_13_PhysRevE.97.062108}. A number of techniques have been developed in recent times to tackle the dynamics of such systems. These include the Hamiltonian of mean force~\cite{Hanggi_RevModPhys.92.041002, Campisi_2009}, reaction coordinate method, which probes strong coupling effects by reaction coordinate mapping of the Hamiltonian~\cite{Anupam_garg, Nazir_2014,nazir_segal_PRXQuantum.4.020307, Hughes_reaction_coordinate}, pseudomodes technique~\cite{Garraway_1996, Garraway_2003}, and the method of hierarchical equations of motion, a numerical approach making use of the influence-functional formalism~\cite{Kubo_tanimura_doi:10.1143/JPSJ.58.101, Tanimura_2016}. In the context of the impact of open system ideas on quantum thermodynamics, the information-theoretic approach provides a deep understanding~\cite{landi_RevModPhys.93.035008}. 

One of the cornerstones of the field of strong-coupling non-Markovian quantum thermodynamics is to correctly define the fundamental thermodynamic quantities such as heat, work, internal energy, ergotropy, and entropy production in the quantum regime~\cite{subotnik_2018, Strasbegr_2019, Rivas_strong_coupling, full_counting_paper, Zhang_2021, Strasberg_Esposito_2017}. This is essential for the definition of thermodynamic laws of strongly coupled NM quantum systems. Further, at strong system and bath couplings, the equilibrium state of the system coupled to a bath has been found to be set by the Hamiltonian of mean force (HMF) instead of the bare Hamiltonian of the system~\cite{miller2018hamiltonian, Hanggi_RevModPhys.92.041002, timofeev2022hamiltonian}. Moreover, a recently developed technique of canonical Hamiltonian and minimal dissipator constructed with traceless Lindblad operators~\cite{Hayden_2022} has been utilized to calculate the thermodynamic quantities of the system~\cite{breuer_effective_Ham1, Colla_Breuer_otto_PhysRevResearch.6.013258}, where the canonical Hamiltonian rather than the system's bare Hamiltonian is used to study the impact of strong coupling and non-Markovian effects on the system. 

In this work, we study the strong-coupling non-Markovian quantum thermodynamics of a central spin interacting with a spin bath~\cite{24t50_38_PhysRevA.96.052125, Devvrat_central_spin1, Bhattacharya2021, devvrat_central_spin_2, Lidar_spin_bath}. The baths can be broadly placed into the following categories~\cite{Banerjee_2007}: (a) Bosonic, (b) fermionic, and (c) spin baths. The spin-boson model~\cite{7t26_1} and the Caldeira-Leggett model~\cite{7t26_4_CALDEIRA1983374} are well-known examples of bosonic baths. The dynamics of these types of models have been studied in the literature~\cite{7t26_6_weiss2012quantum, BreurPoqs, banerjee2018open}. Conducting electrons in the Anderson and the Kondo models are examples of a fermionic bath system. Numerically exact methods, for example, the fermionic hierarchical equation of motion, have been used to obtain the dynamics of the above systems~\cite{fermionic_bath1, fermionic_bath2}. Further, the exact dynamics of noninteracting fermions (bosons) linearly coupled to thermal environments of noninteracting fermions can be found in~\cite{fermionic_bath3}.
Conversely, when dealing with spin baths~\cite{prokofev_stamp_2000}, frequent use is made of perturbative methods or time-dependent quantum master equations~\cite{7t26_18_PhysRevB.70.045323,24t50_51_PhysRevA.76.052119, samya_da_exact_solution}. The dissipative equation of motion approach has also been used to find the exact dynamics of the systems surrounded by a spin bath~\cite{Ying_spin_bath_exact}. Studying these systems is crucial for understanding magnetic systems~\cite{parkinson}, quantum spin glasses~\cite{Rosenbaum}, quantum batteries~\cite{central_spin_quantum_battery}, NV-center~\cite{Hanson_spin_bath}, and superconducting systems~\cite{7t26_18_PhysRevB.70.045323}, among others. Moreover, the finite nature of the bath renders the exact dynamics of the central spin model analytically tractable across arbitrary system-bath coupling regimes. This unique feature provides profound advantages for probing the non-Markovian effects on quantum thermodynamics, particularly in strong coupling regimes.

Here, we develop the exact dynamics of the central spin model. This is then used to understand the quantum thermodynamic properties of the central spin. To this effect, we calculate the thermodynamic quantities of this system using a number of methods outlined above. We begin by analyzing entropy production, heat, and work while discussing the first and second laws of thermodynamics. We also find out the cost of turning the system and bath interactions on or off, and we explore a variation of the second law in finite bath models. Next, we calculate the thermal equilibrium state and entropy under strong coupling using the HMF technique, which aids in examining the third law of quantum thermodynamics. We then discuss the use of the canonical Hamiltonian as a replacement for the system's bare Hamiltonian to calculate the thermodynamic quantities. We aim to arrive at the correct forms of the quantities, such as heat, work, internal energy, entropy production, and thermal equilibrium state in strongly coupled non-Markovian quantum systems and the correct forms of the corresponding laws of quantum thermodynamics, taking the central spin model as an example. Finally, we demonstrate the central spin model as a quantum battery, highlighting how a finite spin bath uniquely charges the central spin quantum battery. To summarize, we first present a brief discussion of the tools for handling strong-coupling non-Markovian quantum thermodynamics (Sec.~\ref{sec_tools_q_thermo}). This is followed in the later sections with new results on the exact dynamics of the central spin model and its quantum thermodynamics in the strong coupling non-Markovian regime, by applying the above tools.

The organization of this work is as follows: In Sec.~\ref{sec_tools_q_thermo}, we discuss tools for exploring non-Markovian strong-coupling quantum thermodynamics. This includes discussions on entropy production, heat, work, the Hamiltonian of mean force, and the canonical Hamiltonian technique. Section~\ref{sec:the_central_spin_model} covers the exact dynamics of the central spin model. In Section~\ref{sec_exploring_NM_q_thermo_cs}, we analyze thermodynamic quantities in the strong coupling and non-Markovian regime of the central spin model. The application of the central spin model as a quantum battery, along with the calculation of its ergotropy, is presented in Section~\ref{sec_ergotropy}. Finally, we conclude in Section~\ref{sec:conclusion}.

\section{Tools for exploring non-Markovian strong-coupling quantum thermodynamics}\label{sec_tools_q_thermo}
Here, we briefly dive into the key tools for probing quantum thermodynamics in a strongly coupled non-Markovian quantum system. 
We initiate the discussion from entropy production for a general quantum system surrounded by a bath (using this, we discuss the second law of quantum thermodynamics), followed by the discussion of the change in the system's and bath's internal energy and the work done as the mismatch between them. This leads us to the first law of quantum thermodynamics in a general system-bath scenario. In the next step, we discuss how the work done is related to the cost of turning the system-bath interactions on or off. To analyze the thermal equilibrium between the system and the bath in the strong-coupling regime, we discuss the formalism of the Hamiltonian of mean force and the mean force Gibbs state. Further, we discuss the canonical Hamiltonian of a system.

Throughout the paper, we consider the evolution of the total system \textit{SB} (system $S$ plus bath $B$) is governed by the Hamiltonian
\begin{align}
    H = H_S + H_B + V,
    \label{total_hamiltonian}
\end{align}
where $H_S$, $H_B$, and $V$ are the system, bath, and system-bath interaction Hamiltonians, respectively. 
The dynamics of the total system \textit{SB} is dictated by the unitary $U = e^{-iHt}$ (for $\hbar = 1$) as 
\begin{align}
    \rho_{SB}(t) = U \left(\rho_S(0) \otimes\rho_B(0)\right) U^\dagger,
    \label{global_dynamics}
\end{align}
where $\rho_S(0)$ and $\rho_B(0)$ are the system's and bath's initial states, and $\rho_{SB}(t)$ is the joint system-bath state at any time $t$. The reduced state of the system ($\rho_S$) or the bath ($\rho_B$) is given by 
\begin{align}
    \rho_{S(B)} = {\rm Tr}_{B(S)}\left[U \left(\rho_S(0) \otimes\rho_B(0)\right) U^\dagger\right].
    \label{eq_reduced_dynamics}
\end{align}

\subsection{Entropy production, heat and work} \label{sec_true_entropy_prodcution}
Considering the global system-bath dynamics for arbitrary initial states of the system and the bath [Eq. (\ref{global_dynamics})], a formulation for the entropy production was developed in \cite{Esposito_2010,landi_RevModPhys.93.035008}. Here, the dynamical evolution of the system is obtained by tracing out the bath, which is the origin of the irreversibility. The entropy production accounts for the contribution of two processes, one where we discard any information stored locally in the state of the bath, and the other where we discard the non-local information shared between the system and the bath, and is given by
\begin{align}
    \Sigma = \mathcal{I}_{\rho_{SB}(t)}(S:B) + S\left[\rho_B(t)\|\rho_B(0)\right],
\end{align}
where $\mathcal{I}_{\rho_{SB}}(S:B) = S(\rho_S) + S(\rho_B) - S(\rho_{SB})$ is the mutual information of any bipartite system $SB$ with $S(\rho) = -{\rm Tr}(\rho\ln\rho)$ being the von Neumann entropy and $S(\rho\|\sigma) = {\rm Tr}\left\{\rho\ln\rho - \rho\ln\sigma\right\}$ is the quantum relative entropy. The first term in the above equation quantifies the amount of shared information that is lost if we trace out the bath, and the second term quantifies how the bath is pushed away from equilibrium. Using the forms of mutual information and quantum relative entropy, the entropy production can be rewritten as
\begin{align}\label{entropy_production}
    \Sigma = S\left[\rho_{SB}(t)\|\rho_S(t)\otimes\rho_B(0)\right].
\end{align}
Further, when we consider the thermal state to be the initial state of the bath ($\rho_B(0) = e^{-\beta H_B}/{\rm Tr}\left[e^{-\beta H_B}\right]$, where $\beta = 1/k_B T$ is the inverse temperature of the bath), the entropy production boils down to 
\begin{align}\label{entropy_prodcution_landi_thermal}
    \Sigma = \Delta S_S + \beta Q_B,
\end{align}
where $\Delta S_S = S[\rho_S(t)] - S[\rho_S(0)]$ is the change in the von Neumann entropy of the system and 
\begin{align}\label{Q_B}
    Q_B = {\rm Tr}\left\{H_B\left[\rho_B(t) - \rho_B(0)\right]\right\}
\end{align}
is the total change in energy of the environment during the unitary evolution of the total system, where $\rho_B(t) = {\rm Tr}_S [\rho_{SB}(t)]$. Equation (\ref{entropy_prodcution_landi_thermal}) with $\Sigma \ge 0$ is the standard form of the second law of thermodynamics, and it is in the form of Clausius' inequality~\cite{Clausius1865}. The derivation of Eq.~\eqref{entropy_prodcution_landi_thermal} from Eq.~\eqref{entropy_production} can be done without taking any assumptions about the size of the bath. However, for a finite bath, an expression for the entropy production $\Sigma_{\rm finite} = \Delta S_S + \int \delta Q_B(t)/T(t) \ge 0$ was suggested in~\cite{Strasberg2021, Strasberg_2021_2}, where the temperature $T(t)$ of the finite bath changes with time. Note that if the bath is large enough, such that its temperature stays constant, $\Sigma_{\rm finite}$ reduces to $\Sigma$. Further, although $\Sigma \ge 0$ is still a valid inequality, it was argued that the second law should be identified by $\Sigma_{\rm finite}\ge 0$~\cite{Strasberg2021} for a finite bath. Moreover, $\Sigma_{\rm finite}$ is related to $\Sigma$ by~\cite{Strasberg2021}
\begin{align}
    \Sigma_{\rm finite} = \Sigma - S\left(\zeta_B\left[T(t)\right]||\zeta_B\left[T(0)\right]\right),
    \label{eq_finite_entropy_production}
\end{align}
where $\zeta_B\left[T(t)\right] = e^{-H_B/T(t)}/{\rm Tr}\left[e^{-H_B/T(t)}\right]$. In the finite spin bath system studied here, we assume that all bath spins are initially in thermal equilibrium at a temperature $T(0) = T$. As the system evolves, the bath's temperature changes due to its finite size and its interaction with the central spin.

Next, the change in the internal energy of the system is given by 
\begin{align}
    \Delta U_S = {\rm Tr}\left\{\rho_S(t) H_S\right\} - {\rm Tr}\left\{\rho_S(0) H_S\right\}.
\end{align}
From the first law of thermodynamics, the total change in the internal energy of the system is given by 
\begin{align}\label{eq_first_law}
    \Delta U_S = W - Q_B,
\end{align}
where $W$ is the work performed by the bath, with $W>0$ means the work is performed on the system and $Q_B$ is given in Eq. (\ref{Q_B}). Meanwhile, the work ($W$) can also be seen as the mismatch between the energy changes of the system $\Delta U_S$ and the bath $Q_B$. 

The change in the internal energy of the bath can also be calculated using the bath heat current~\cite{Tanimura_2016}. Consider a system $H_S'$ coupled to a heat bath $H_B'$ via an interaction Hamiltonian $H_{I}'$. The system heat current $\dot Q_S'(t)$, in this case, is given by $\dot Q_S'(t) = \frac{d}{dt}\braket{H_S'(t)} - \dot W'(t)$, where $\dot W'(t) = \left\langle\frac{\partial H_S'(t)}{\partial t}\right\rangle$. The average $\left\langle\cdot\right\rangle = {\rm Tr}\left[\rho_{SB}(t)\cdot\right]$. Using the relation between the expectation value of a quantum mechanical operator and the expectation value of the commutator between the operator and the Hamiltonian of the system, the system heat current can be rewritten as
\begin{align}
    \dot Q_S'(t) = i\left\langle[H_I'(t), H_S'(t)]\right\rangle,
\end{align}
which for a constant $H_S'$ becomes the rate of change in the system's internal energy. 
Considering the rate of change in the bath energy $\dot Q_B'(t) = \frac{d}{dt}\left\langle H_B'(t)\right\rangle$ for a constant $H_B'$, we get the following relation between the $\dot Q_B'(t)$ and the $\dot Q_S'(t)$
\begin{align}
    \dot Q_B' (t) = -\dot Q_S'(t) - \frac{d}{dt}\braket{H_I'(t)}.  
\end{align}
Replacing $H_I'$ and $H_S'$ in the above equation with the interaction Hamiltonian $V$ and the system Hamiltonian $H_S$, respectively, from Eq. (\ref{Hname}), we get an alternate expression for the change in the internal energy of the bath $Q_B$, that is, 
\begin{align}\label{bath_heat_current_bare}
    Q_B &= \int_0^t d\tau \left\{-\dot Q_S(\tau) - \frac{d}{d\tau}\braket{V(\tau)}\right\}\nonumber\\
    &= -\int_0^t d\tau \left\{i\left\langle[V(\tau), H_S(\tau)]\right\rangle\right\} + \braket{V(0)} - \braket{V(t)}.
\end{align}
Notice that the first term on the right side of the above equation denotes the negative of the change in the internal energy of the system for a constant $H_S$, which is the case considered here. To this end, the second term on the RHS of the above equation balances the difference between the change in the internal energies of the system and the bath. Therefore, this term provides the work done by the bath on the system, that is, 
\begin{align}\label{work_done_by_bath}
    W &= {\rm Tr}\left[V\left\{\rho_{SB}(0) - \rho_{SB}(t)\right\}\right].
\end{align}
This form is particularly useful as it brings out how, in a general scenario of strong system-bath coupling and under constant Hamiltonians, the work done by the bath can be explicitly computed, saving the need to calculate the change in the internal energies of the system and the bath. It is also worth mentioning that this amount of work must be supplied to turn off (or on) the interaction between the system and the bath. 

\subsection{The Hamiltonian of mean force}
Consider a quantum system $S$ interacting with a thermal bath $B$, which is described by the Hamiltonian $H$ in Eq.~\eqref{total_hamiltonian}, 
The global equilibrium state of the system and the reservoir at temperature $T$ is denoted by ${\zeta}_{SB}$, which is of Gibbs form,
\begin{equation}
    {\zeta}_{SB} := \frac{\exp(-\beta H)}{{Z}_{SB}},
\end{equation}
where, ${Z}_{SB} = {\rm Tr} \left[\exp(-\beta {H})\right]$ is the partition function for $SB$, and $\beta = (k_{B}T)^{-1}$ ($k_B$ is chosen to be one throughout the paper). The interaction term causes the reduced equilibrium state of the system $S$, ${\zeta}^*_{S}(T) = {\rm Tr}_{B} \left[{\zeta}_{SB}(T)\right] $, to deviate from the thermal state ${\zeta}_{S}(T) = \exp{-\beta H_S}/{\rm Tr}\left[\exp{-\beta H_S}\right]$ unless the system-bath coupling is very weak.
The reduced equilibrium state of the system $S$ as an effective Gibbs state is then given by ${\zeta}^*_{S}(T) = \frac{\exp\left[-\beta {\mathcal{H}}_{S}^* (T)\right]}{{Z}_{S}^*}$, where the partition function for the system S can be expressed as the ratio 
\begin{equation}
    {Z}_S^{*} = {Z}_{SB}/{Z}_{B},
\end{equation} where ${Z}_{SB}= {\rm Tr}[e^{-\beta {{H}}]}$ and ${Z}_{B}= {\rm Tr}_{B}[e^{-\beta {H}_{B} }]$ and,
\begin{equation}
    \mathcal{{H}}^*_S (T) := - \frac{1}{\beta} \log \left( \frac{ {\rm Tr}_{B} \left[\exp(-\beta {{H}}) \right]}{ {\rm Tr}_{B} \left[\exp(-\beta {H_{B}})\right]}\right),
    \label{eq_Hamiltonian_of_mean_force}
\end{equation}
is the Hamiltonian of mean force~\cite{kirkwood1935statistical, timofeev2022hamiltonian, miller2018hamiltonian}. 
This operator acts as an effective Hamiltonian for the system \( S \), incorporating both temperature \( T \) and interaction \( V \). In the weak coupling regime, it simplifies to the bare Hamiltonian \( H_S \). The Hamiltonian of mean force (HMF) approach is crucial for understanding key thermodynamic properties like free energy, entropy, and heat capacities in systems with significant environmental interactions~\cite{miller2018, pathania2024}. It aids in formulating generalized thermodynamic potentials and investigating equilibrium and non-equilibrium processes, shedding light on quantum dissipation and decoherence.
Further, for the thermal equilibrium state $\zeta_S (T)$, the entropy of the system is given by $\mathcal{S_S} = -{\rm Tr}\left[\zeta_S(T) \log\zeta_S(T)\right]$ (von Neumann entropy) in the weak coupling regime. In the strong coupling regime due to the alternative partition function ${Z}_S^{*}$, the system's thermodynamic entropy becomes~\cite{miller2018hamiltonian}
\begin{align}
    \mathcal{S}_S = -{\rm Tr}\left[\zeta^*_S(T) \log\zeta^*_S(T)\right] + \beta^2 {\rm Tr}\left[\partial_\beta\mathcal{H}_S^*(T)\zeta_S^*(T)\right].
    \label{thermodynamic_entropy}
\end{align}
The first term is the von Neumann entropy of the system, while the second term comes from the temperature dependence of HMF. This suggests that thermodynamic entropy is not universally equivalent to the information content of the equilibrium state in systems with interactions.   

\subsection{The theory of canonical Hamiltonian}\label{sec:theory_of_can_ham}
The evolution of a closed quantum system under the influence of a Hamiltonian $H$ is obtained using the von Neumann equation
\begin{equation}
    \frac{d \rho}{d t}=-i[H,\rho],
\end{equation}
where $\rho$ is the density matrix of the closed quantum system. 
The equation of motion assumes an even more sophisticated form when the closed system is broken into two subsystems, say $S$ and $B$. In general, $S$ is referred to as an open quantum system (whose dynamics is of interest) surrounded by a bath $B$. In this scenario, the Hilbert space structure becomes $\mathcal{H}=\mathcal{H}_S \otimes \mathcal{H}_B$ and the dynamics  of subsystem $S$ boils down to
\begin{equation}
    \frac{d \rho_S}{d t}=\mathcal{L}_t (\rho_S),
\end{equation}
where $\mathcal{L}_t$ is a linear superoperator in a finite-dimensional Hilbert space of $\mathcal{H}_S$. 
This $\emph{Hermiticity-preserving}$ and $\emph{Trace Annihilating}$ or HPTA dynamical map $\mathcal{L}_t$ can be redefined if a suitable yet arbitrary Hermitian operator $H_S(t)$ is chosen. To this end, the new form becomes
\begin{equation}
    \frac{d \rho_S}{d t}=-i[H_S(t),\rho_S] + \mathcal{D}_t(\rho_S).
\end{equation}
This relatively convenient separation of terms involves one $\emph{Hamiltonian}$ part $H_S(t)$ and one $\emph{dissipator}$ part $\mathcal{D}_t$, which collectively puts it into the form of a $\emph{quantum master equation}$. From \cite{7t26_3_lindblad1976generators,ECG_10.1063/1.522979}, it can be shown that there always exists jump operators $L_j$ and real coefficients $\gamma_j$ that puts the dissipator part into the form, 
\begin{equation}\label{eq:dissipator_form}
    \mathcal{D}(\rho_S)=\sum_j \gamma_j \bigg( L_j\rho_S L_j^{\dagger}-\frac{1}{2}\bigg\{ L_j^{\dagger}L_j, \rho_S \bigg\} \bigg).
\end{equation}
Even in this `Lindbladian' form, there are ambiguities. Consider the set of transformations, \cite{Hayden_2022,BreurPoqs}
\begin{align}
    L_j &\rightarrow L_j + \alpha_j(t)I_S,\\
    H_S(t) &\rightarrow H_S(t) + \sum_j \frac{\gamma_j}{2i} \big( \alpha_j L_j^{\dagger}-\bar{\alpha}L_j \big),
\end{align}
on the Lindblad (jump) and Hamiltonian operators, respectively. It leaves the structure $-i[H_S(t),\rho_S]+\mathcal{D}_t(\rho_S)$ intact.
To get rid of this ambiguity and to render a unique dissipator, the jump operators $L_j'$s need to be traceless, as shown in \cite{ECG_10.1063/1.522979}.
When the Lindbladian dissipator is so structured, then the Hamiltonian part of the master equation becomes the canonical Hamiltonian. This canonical Hamiltonian contains information beyond just the system Hamiltonian about the part of the open system that evolves unitarily. 
A convenient way to find out the canonical Hamiltonian is given by
\begin{align}
    H^{\rm can}_S&=\frac{1}{2id}\sum_j \gamma_j \bigg({\rm Tr}(E_j)E_j^{\dagger}-{\rm Tr}(E_j^{\dagger})E_j\bigg),
\end{align}
where the $E_j$'s are the pseudo-Kraus operators obtained from the decomposition of the Lindbladian superoperator and $d$ is the dimension of the system~\cite{Hayden_2022}.
Just like any completely positive superoperator admits a Kraus operator representation, any HPTA superoperator admits a pseudo-Kraus representation, 
\begin{equation}
    \mathcal{L}(\rho)=\sum_j \gamma_j E_j \rho E_j^{\dagger},
\end{equation}
where $\gamma_j$'s are real coefficients. Further, the form of the minimal dissipator is given by 
\begin{align}\label{eq:minimal_dissipator}
    \mathcal{D}_t(\rho) &= \mathcal{L}(\rho) + i [H_S^{\rm can}, \rho]\nonumber \\
    &= \sum_j\gamma_j \left(E_j\rho E_j^\dagger + \frac{1}{2d}\left[{\rm Tr}(E_j)E_j^\dagger - {\rm Tr}(E_j^\dagger)E_j, \rho\right]\right), 
\end{align}
which upon using the identity $\sum_j\gamma_jE_j^\dagger E_j = 0$ reduces to the form of Eq.~\eqref{eq:dissipator_form}, with $L_j$ being given by 
\begin{align}
    L_j = E_j - \frac{{\rm Tr}(E_j)}{d}\mathbb{I}.
\end{align}

\section{The central spin model and its exact dynamics}\label{sec:the_central_spin_model}
In the central spin model, a spin-$1/2$ particle is surrounded by $N$ other spins in a circular formation~\cite{Bhattacharya2021, Devvrat_central_spin1, devvrat_central_spin_2}, and the central spin interacts uniformly with the bath spins. Here, the total Hilbert space of all $N$ bath spins is conveniently reduced to an $N + 1$ dimensional space using the collective angular momentum operators. The Hamiltonian (for $\hbar = 1$) of the total system is given by,
\begin{align}\label{Hname}
    H &= H_S + H_B + V \nonumber \\ 
    &=\frac{ \omega_0}{2}\sigma_z^0 + \frac{ \omega}{N} J_z + \frac{\epsilon}{\sqrt{N}} \left(\sigma^0_xJ_x+\sigma^0_y J_y\right),
\end{align}
where $\omega_0$ is the transition frequency of the central spin-$1/2$ particle, $\epsilon$ is the interaction strength between the system and the bath, with $\sqrt{N}$ being the scaling factor. $\omega/N$ is the scaled frequency of the bath. The evolution of the composite system \textit{SB} for an arbitrary initial state $\rho_{SB}(0) = \rho_S(0) \otimes \rho_B(0)$, assuming separable initial condition, of the total system (where $\rho_S(0)$ and $\rho_B(0)$ are the initial states of the central spin and the bath, respectively) by means of the global unitary $U = e^{-iHt}$ is given by 
\begin{align}\label{eq_total_unitary_evolution}
    \rho_{SB}(t) = e^{-iHt}\left\{\rho_S(0) \otimes\rho_B(0)\right\}e^{iHt}.
\end{align}
The reduced state of the central spin system after tracing out the bath is given by
\begin{align}
    \rho_S(t)={\rm Tr}_B\left[ e^{-iHt}\left\{\rho_S(0) \otimes \rho_B(0) \right\}e^{iHt}\right],
    \label{Eq_unitary_dynamics}
\end{align}
where $H$ is the total Hamiltonian. Here, we consider the Gibbs state as the initial state of the bath, which is given (using the spectral decomposition of the bath Hamiltonian $H_B = \frac{\omega}{N}J_z = \frac{\omega}{N}\sum_{n=0}^N\left\{\left(\frac{N}{2} - n\right)\ket{n}\bra{n}\right\} = \frac{\omega}{2}\sum_{n=0}^N\left\{\left(1 - \frac{2n}{N}\right)\ket{n}\bra{n}\right\}$) by
\begin{align}\label{rhoB0}
    \rho_B(0)&= \frac{e^{-\beta H_B}}{Z}  =\frac{1}{Z}\sum_{n=0}^{N}e^{-\frac{\beta\omega}{2}\left(1-\frac{2n}{N}\right)}\ket{n}\bra{n},
\end{align}
where $\beta = 1/T$, $\ket{n}$ is the standard computational basis, and $Z = \sum_{n=0}^{N}e^{-\frac{\beta\omega}{2}\left(1-\frac{2n}{N}\right)}$. 
To find out the exact dynamics of the system, we derive the spectral decomposition of the total Hamiltonian $H$. The nature of the eigenvalues and eigenvectors of the total Hamiltonian of the central spin model is illustrated in Appendix A. We use these eigenvalues and eigenvectors in Eq. (\ref{Eq_unitary_dynamics}) to obtain the exact dynamics of the central spin system. Considering an arbitrary initial state of the system $\rho_S(0) = \begin{pmatrix}
    \rho_{00}(0) && \rho_{01}(0)\\\rho_{10}(0) && \rho_{11}(0)
\end{pmatrix}$, the density matrix dictating the evolution of the central spin is given by
\begin{widetext}
\begin{align}\label{rhos(t)}
    \rho_S(t)=\begin{pmatrix}
        \alpha(t)\rho_{00}(0)+\eta(t)\rho_{11}(0) && \delta(t)\rho_{01}(0) \\
        \delta^*(t)\rho_{10}(0) &&  (1-\alpha(t))\rho_{00}(0)+(1-\eta(t))\rho_{11}(0)     
    \end{pmatrix},
\end{align}
where $\alpha(t)=\frac{1}{Z}\left[ e^{-\frac{\beta\omega}{2}} + S_n(t) \right]$ with 
\begin{align}
    S_n(t)&=\sum^N_{n=1}e^{-\frac{\beta\omega}{2}\left(1-\frac{2n}{N}\right)} \left\{ \left(\frac{\chi_+(n)^2}{1+\chi_+(n)^2}\right)^2 +\left(\frac{\chi_-(n)^2}{1+\chi_-(n)^2}\right)^2+\left(\frac{\chi_+(n)^2 \chi_-(n)^2}{\left[1+\chi_+(n)^2\right]\left[1+\chi_-(n)^2\right]}\right)2\cos\left(\frac{t\sqrt{b^2+4a_n^2}}{N}\right)\right\}. 
\end{align}
Further, 
\begin{align}\label{eq_etat}
\eta(t)&=\frac{1}{Z}\sum^N_{n=1}e^{-\frac{\beta\omega}{2}\left(1-\frac{2(n-1)}{N}\right)} \left\{ \left(\frac{\chi_+(n)}{1+\chi_+(n)^2}\right)^2 +\left(\frac{\chi_-(n)}{1+\chi_-(n)^2}\right)^2+\left(\frac{\chi_+(n) \chi_-(n)}{\left[1+\chi_+(n)^2\right]\left[1+\chi_-(n)^2\right]}\right) 2\cos\left(\frac{t\sqrt{b^2+4a_n^2}}{N}\right)\right\},
\end{align}
and in the non-diagonal terms,
\begin{align}
    \delta^*(t)&=\frac{1}{Z}\left\{\frac{1}{1+\chi_+(1)^2} \left[ e^{-i\left\{\lambda_+(1)-\frac{(\omega +\omega_0)}{2}\right\}t-\frac{\beta\omega}{2}} + \chi_+(N)^2 e^{i\left\{\lambda_+(N)-\frac{(\omega +\omega_0)}{2}\right\}t+\frac{\beta\omega}{2}}  \right] + \frac{1}{1+\chi_-(1)^2}\left[ e^{-i\left\{\lambda_-(1)-\frac{(\omega +\omega_0)}{2}\right\}t-\frac{\beta\omega}{2}} + \chi_-(N)^2 e^{i\left\{\lambda_-(N)-\frac{(\omega +\omega_0)}{2}\right\}t+\frac{\beta\omega}{2}} \right]\right\} \nonumber \\
    &+\frac{1}{Z} \sum^N_{n=2}e^{-\frac{\beta\omega}{2}\left[1-\frac{2(n-1)}{N}\right]}\left\{ \left(\frac{e^{-i\lambda_+(n)t}}{1+\chi_+(n)^2}+\frac{e^{-i\lambda_-(n)t}}{1+\chi_-(n)^2}\right) \times \left(e^{i\lambda_+(n-1)t}\frac{\chi_+(n-1)^2}{1+\chi_+(n-1)^2}+e^{i\lambda_-(n-1)t}\frac{\chi_-(n-1)^2}{1+\chi_-(n-1)^2}\right)\right\}.
\end{align}
\end{widetext}
Here, $\chi_\pm(n), b$ and $a_n$ come from the structure of the eigenvectors of the total Hamiltonian $H$ discussed in Appendix A.
The dynamical map $\Phi(t)$ of the system in the superoperator space is given by 
\begin{align}\label{lambda and chi struc}
    \Phi(t)=\begin{pmatrix}
        \alpha(t) & 0 & 0 & \eta(t)\\
        0 & \delta(t) & 0 & 0 \\
        0 & 0 & \delta^*(t) & 0\\
        1-\alpha(t) & 0 & 0 & 1-\eta(t)
    \end{pmatrix}.
\end{align}
Now, from the relation $\mathcal{L}=\Dot{\Phi(t)}\Phi(t)^{-1}$, one can find out the matrix form of the Lindbladian superoperator, given below 
\begin{align}
    \mathcal{L}=\begin{pmatrix}
        \frac{\Dot{\alpha}(t)[1-\eta(t)]+\Dot{\eta}(t)[\alpha(t)-1]}{\alpha(t)-\eta(t)} & 0 & 0 & \frac{-\Dot{\alpha}(t)\eta(t)+\Dot{\eta}(t)\alpha(t)}{\alpha(t)-\eta(t)}\\
        0 & \frac{\Dot{\delta}(t)}{\delta(t)} & 0 & 0 \\
        0 & 0 & \frac{\Dot{\delta}^*(t)}{\delta^*(t)} & 0\\
        \frac{-\Dot{\alpha}(t)[1-\eta(t)]-\Dot{\eta}(t)[\alpha(t)-1]}{\alpha(t)-\eta(t)} & 0 & 0 & \frac{\Dot{\alpha}(t)\eta(t)-\Dot{\eta}(t)\alpha(t)}{\alpha(t)-\eta(t)}
    \end{pmatrix}. 
\end{align}%

\begin{figure*}
    \centering
    \includegraphics[width=1.75\columnwidth]{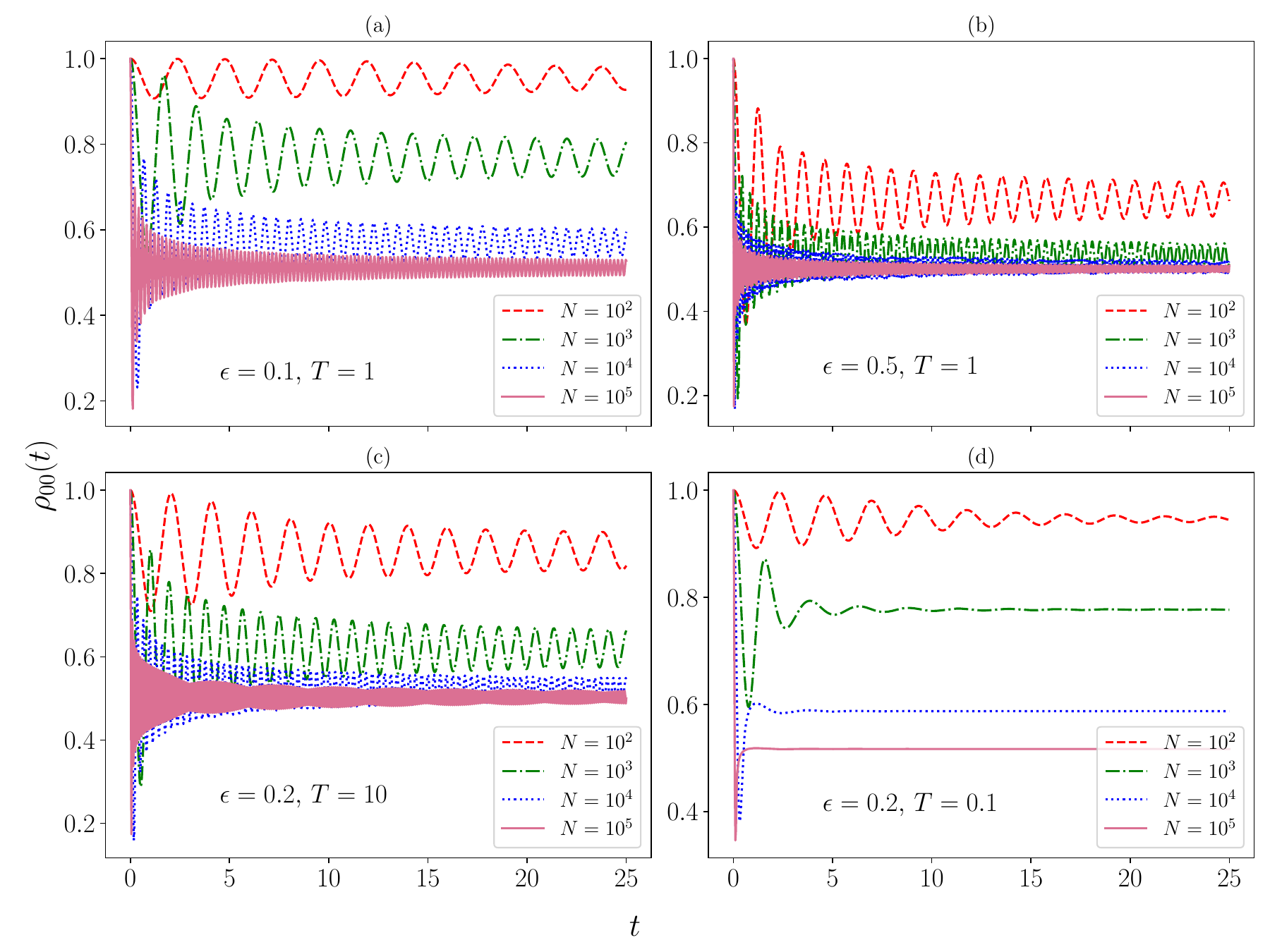}
    \caption{Variation of the central spin system's density matrix element $\rho_{00}(t)$ with time for different values of the number of spins $N$ in the spin bath. Subplot (a) has $\epsilon = 0.1, T = 1$, (b) has $\epsilon = 0.5, T = 1$, (c) has $\epsilon = 0.2, T = 10$, and (d) has $\epsilon = 0.2, T = 0.1$. Further, $\omega_0 = 2.5$ and $\omega = 2.0$ for all the plots. The system is initially taken to be in the excited state ($\ket{\psi(0)}_S = \ket{0}_S$).}
    \label{fig_large_N_limit}
\end{figure*}%
Before exploring the quantum thermodynamic properties of the central spin system, we examine the impact of the number of bath spins $N$, temperature $T$, and coupling strength $\epsilon$ on the system's dynamics, as can be seen from Fig.~\ref{fig_large_N_limit}. We plot the variation in the excited state population $\rho_{00}(t)$ of the system's state with time for different values of $\epsilon$ and $T$ in the large $N$ limit. $\rho_{00}(t)$ plays an important role in the system's evolution and is closely connected to the change in its internal energy, see Eq.~\eqref{Delta_US} below. In fact, $\Delta U_S(t) = \omega_0\left[\rho_{00}(t)-\rho_{00}(0)\right]$, which for initial excited state becomes $\omega_0\left[\rho_{00}(t)-1\right]$. From Fig.~\ref{fig_large_N_limit}, we observe that as the number of bath spins increases from $N = 10^2$ to $10^5$, approaching the thermodynamic limit, the frequency of oscillations of the excited state population rapidly increases. However, this increase in oscillation frequency is accompanied by a reduction in amplitude over time, particularly for larger $N$. These effects become even more pronounced as the system-bath coupling is increased $\epsilon = 0.5$. Additionally, temperature plays a crucial role in influencing the system's evolution. At low temperatures, the oscillations in the excited state population diminish over time. Notably, in this regime, Fig.~\ref{fig_large_N_limit}(d), the system's steady state is quickly achieved (the oscillations in the excited state population quickly die) as the number of bath spins is increased. This behavior suggests that the coupling strength and the number of bath spins, in conjunction with the bath's temperature, play a fundamental role in dictating the system's approach to equilibrium.

Having discussed the dynamics of the central spin system, we now move on to study its thermodynamic aspects in the subsequent section. 

\begin{figure*}
    \centering
    \includegraphics[width = 1.75\columnwidth]{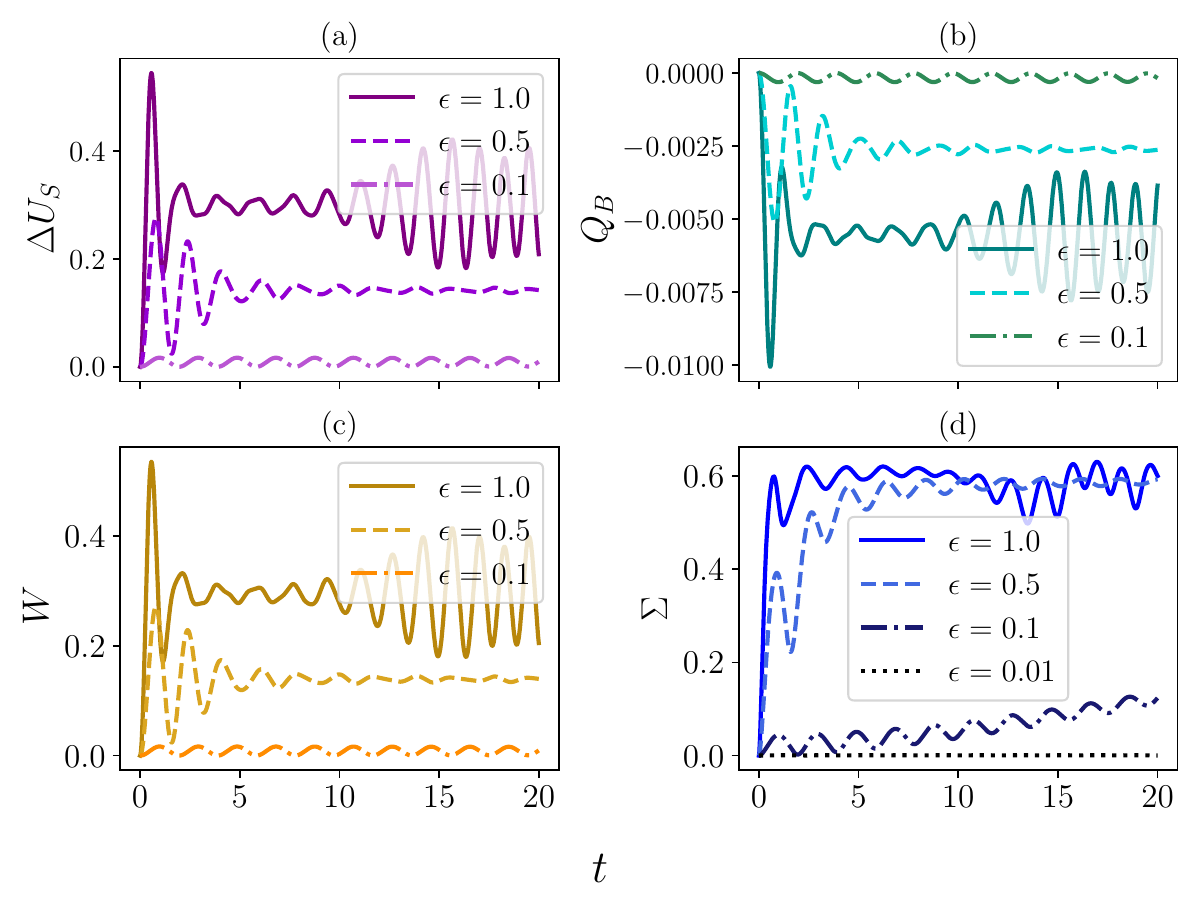}
    \caption{Variation of (a) change in the internal energy of the system $\Delta U_S$, (b) change in the energy of the bath $Q_B$, (c) Work done by the bath $W$, and (d) entropy production $\Sigma$ with time $t$. The parameters are chosen to be $N = 50$, $\omega = 3$, $\omega_0 = 3.25$, and $T = 0.25$. Further, the system's initial state is taken to be $\frac{1}{2}\ket{0} + \frac{\sqrt{3}}{2}\ket{1}$.}
    \label{fig_heat_work_entropy}
\end{figure*}

\section{Exploring the NM quantum thermodynamics of central spin model}\label{sec_exploring_NM_q_thermo_cs}
Here, we calculate the thermodynamic quantities discussed in Sec.~\ref{sec_tools_q_thermo} using the exact dynamics of the central spin model discussed above. 
\subsection{Entropy production, heat, work and the laws of quantum thermodynamics}
We use the Bloch vector representation for the reduced state of the central spin system to calculate the thermodynamic quantities discussed above. The Bloch vector representation of a single qubit density matrix is given by 
\begin{equation}\label{eq_Bloch_vector_form}
    \rho_S(t) = \frac{1}{2}\begin{pmatrix}
        1 + z(t) && x(t) - iy(t)\\
        x(t) + iy(t) && 1 - z(t)
    \end{pmatrix},
\end{equation}
where $k(t) = {\rm Tr}\left\{\sigma_k\rho_S(t)\right\}$ for $(k = x, y, z)$, with $\sigma_k$ being the Pauli spin matrices. On comparing the above equation with Eq. (\ref{rhos(t)}), we get 
\begin{align}\label{xt_yt_zt}
    z(t) &= 2\left[\alpha(t)\rho_{00}(0) + \eta(t) \rho_{11}(0)\right] - 1, \nonumber \\
    x(t) &= 2\Re\left\{\delta (t)\rho_{01}(0)\right\},\nonumber \\
    y(t) & = -2\Im\left\{\delta (t) \rho_{01}(0)\right\}.
\end{align}
The change in the internal energy $\Delta U_S$ of the central spin system is now given by 
\begin{align}\label{Delta_US}
    \Delta U_S = \frac{\omega_0}{2}\left[z(t) - z(0)\right]. 
\end{align}
The calculation of the reduced state of the bath is cumbersome. Hence, we resort to numerical means by tracing out the system from the composite system-bath state [Eq. (\ref{eq_total_unitary_evolution})], that is,
\begin{align}
    \rho_B(t) = {\rm Tr}_S \left[e^{-iHt}\left\{\rho_S(0)\otimes\rho_B(0)\right\}e^{iHt}\right].
\end{align}
This state is used to calculate the change in energy of the bath $Q_B$, and then the work done $W$ by the bath is calculated using $W = \Delta U_S + Q_B$, where $\Delta U_S$ is obtained from Eq. (\ref{Delta_US}).

The variations of the above-mentioned thermodynamic quantities with time are plotted in Fig. \ref{fig_heat_work_entropy}. We observe that the change in the internal energy of the system $\Delta U_S$ is positive and undergoes revivals in time, as depicted by the oscillations. For higher values of the system-bath interaction parameter $\epsilon$, we observe well-defined wavepacket-like structures. 
Further, at a very low value of $\epsilon$, the system's internal energy change is approximately zero. The above observation matches the variations in the change in the energy of the bath $Q_B$ and the work done by the bath $W$. The values of the $Q_B$ are approximately zero for weak coupling ($\epsilon/\omega_0 \approx 0.03$), which is consistent with the GKLS condition, and are otherwise negative, denoting that the energy of the bath decreases during the dynamics, and the positive values of $W$ denote that the bath works on the system. This observation is consistent with the first law of thermodynamics; that is, the energy of the bath decreases while it works on the system. In the process, the internal energy of the system increases. Interestingly, from Fig.~\ref{fig_heat_work_entropy}(c), we observe that the amount of work required to turn off the interaction is approximately zero in the case of weak coupling ($\epsilon/\omega_0 \approx 0.03$) and increases with an increase in the coupling strength. This observation is consistent with the literature~\cite{full_counting_paper}.

Furthermore, we plot the variation of the entropy production $\Sigma$ with time in Fig. \ref{fig_heat_work_entropy}(d). We observe that the entropy production is positive during the system's dynamics, which is consistent with the second law of thermodynamics. Moreover, the entropy production is zero for the very weak interaction case, reminiscent of the GKLS condition, as compared to the strong one. The entropy production for the values of interaction strength $\epsilon$ between 0.5 and 1 saturates around 0.6, with the latter case being oscillatory. Interestingly, the oscillations in the values of entropy production indicate a negative entropy production rate, which is generally observed in non-Markovian systems~\cite{Rivas_strong_coupling, breuer_effective_Ham1}. 

\begin{figure}
    \centering
    \includegraphics[width = 1\columnwidth]{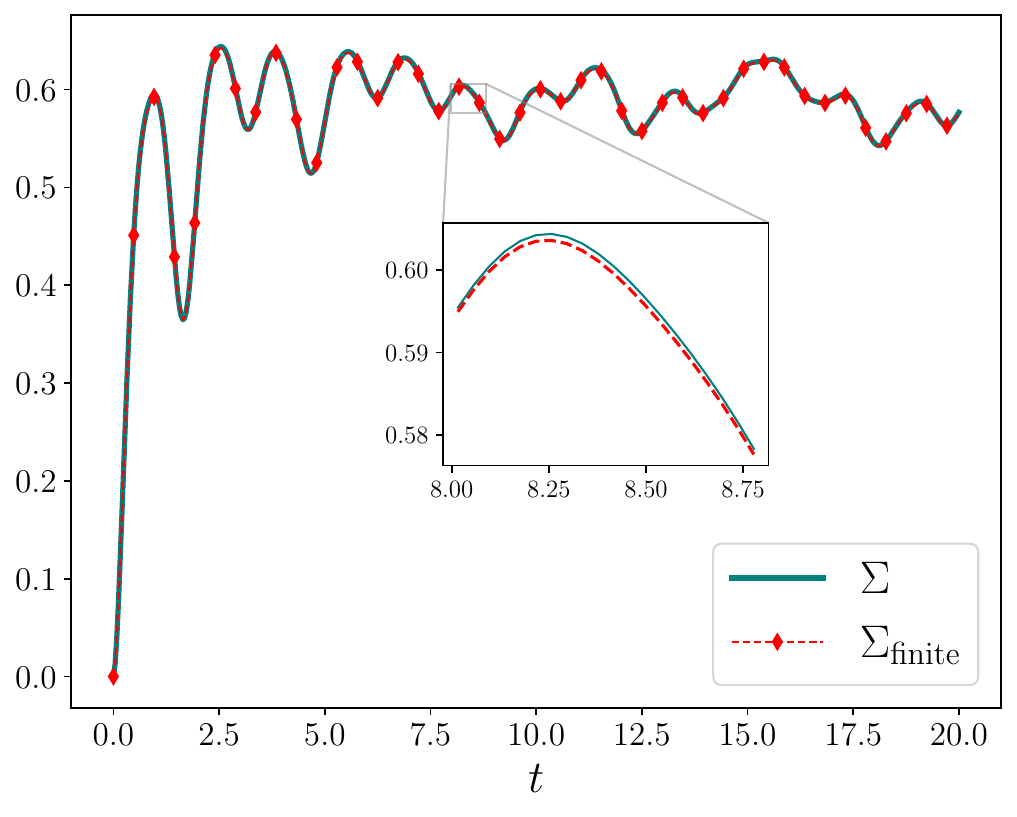}
    \caption{A comparison of the entropy production $\Sigma$ calculated using Eq.~\eqref{entropy_prodcution_landi_thermal} with the entropy production for finite baths $\Sigma_{\rm finite}$ calculated using Eq.~\eqref{eq_finite_entropy_production}. The parameters are chosen to be $N = 10, T = 1, \omega = 2, \epsilon = 1$, and $ \omega_0 = 2.5$.}
    \label{fig_finite_entropy_production}
\end{figure}

Also, we calculate the entropy production explicitly for the finite bath $\Sigma_{\rm finite}$ discussed in Eq.~\eqref{eq_finite_entropy_production}. To find out the temperature variation with time $T(t)$, we make us of the suggestion in~\cite{Strasberg2021} to numerically fit the variable $T$, such that the quantity ${\rm Tr}\left[H_B\zeta_B\{T(t)\}\right]$ becomes equal to the bath's internal energy ${\rm Tr}\left[H_B\rho_B(t)\right]$ at time $t$, Eq.~\eqref{Q_B}. We then use the time-dependent bath temperature $T(t)$ to calculate the quantity $\Sigma_{\rm finite}$. The plot for $\Sigma_{\rm finite}$ is given in Fig.~\ref{fig_finite_entropy_production}. It can be observed that the variation of $\Sigma_{\rm finite}$ is similar to $\Sigma$. The inset plot shows that the entropy production for the finite bath $\Sigma_{\rm finite}$ (red dashed curve) is upper bounded by the quantity $\Sigma$ (teal solid curve), which is consistent with the inequality $\Sigma \ge \Sigma_{\rm finite}$ suggested in~\cite{Strasberg2021}.

\subsection{HMF and thermodynamic entropy of the central spin model}
Here, we use the total Hamiltonian's spectral decomposition (see Appendix A) to calculate the HMF of the central spin model using Eq.~\eqref{eq_Hamiltonian_of_mean_force}. The analytical form of the HMF for the central spin model is given by
\begin{align}
    \mathcal{H}_S^*(\beta) = -\frac{1}{\beta}\begin{pmatrix}
        \log\left[\pi_{00}(\beta)\right] &0 \\
        0& \log\left[\pi_{11}(\beta)\right]
    \end{pmatrix},
\end{align}
where 
\begin{align}
\pi_{00}(\beta) &= \frac{1}{\xi(\beta)}\left[\sum_{n = 1}^N\left\{\frac{e^{-\beta \lambda_\pm(n)}\chi_\pm(n)^2}{1 + \chi_\pm(n)^2}\right\} + e^{-\beta(\omega + \omega_0)/2} \right], \nonumber \\
\pi_{11}(\beta) &= \frac{1}{\xi(\beta)}\left[\sum_{n = 1}^N\left\{\frac{e^{-\beta \lambda_\pm(n)}}{1 + \chi_\pm(n)^2}\right\} + e^{\beta(\omega + \omega_0)/2} \right],
\label{Eq_pi_00_and_pi_11}
\end{align}
and $\xi(\beta) = \frac{\sinh[\beta \omega (N+1)/2N]}{\sinh(\beta\omega/2N)}$. The forms of the factors $\lambda_\pm(n)$ and $\chi_\pm(n)$ are given in Appendix A. We further calculate the thermodynamic entropy of the central spin system using the above HMF in Eq.~\eqref{thermodynamic_entropy}. The analytical form of the thermodynamic entropy is given by 
\begin{align}
    \mathcal{S}_S &= -\frac{\pi_{00}(\beta)}{\pi_{00}(\beta) + \pi_{11}(\beta)}\log\left(\frac{\pi_{00}(\beta)}{\pi_{00}(\beta) + \pi_{11}(\beta)}\right) \nonumber \\
    &- \frac{\pi_{11}(\beta)}{\pi_{00}(\beta) + \pi_{11}(\beta)}\log\left(\frac{\pi_{11}(\beta)}{\pi_{00}(\beta) + \pi_{11}(\beta)}\right) \nonumber \\
    &+ \beta^2 \left(\frac{\mu_{00}(\beta)\pi_{00}(\beta) + \mu_{11}(\beta)\pi_{11}(\beta)}{\pi_{00}(\beta) + \pi_{11}(\beta)}\right), 
    \label{eq_thermodynamic_entropy_cs}
\end{align}
where 
\begin{align}
    \mu_{00}(\beta) &= \frac{\log[\pi_{00}(\beta)]}{\beta^2} - \frac{\partial_\beta \pi_{00}(\beta)}{\beta\pi_{00}(\beta)}, \nonumber \\
    \mu_{11}(\beta) &= \frac{\log[\pi_{11}(\beta)]}{\beta^2} - \frac{\partial_\beta \pi_{11}(\beta)}{\beta\pi_{11}(\beta)}.
\end{align}
The partial differentiation $\partial_\beta \pi_{xx}(\beta)$ for $(x =0, 1)$ can be easily calculated from Eq.~\eqref{Eq_pi_00_and_pi_11}.
\begin{figure*}
    \centering
    \includegraphics[width = 1.75\columnwidth]{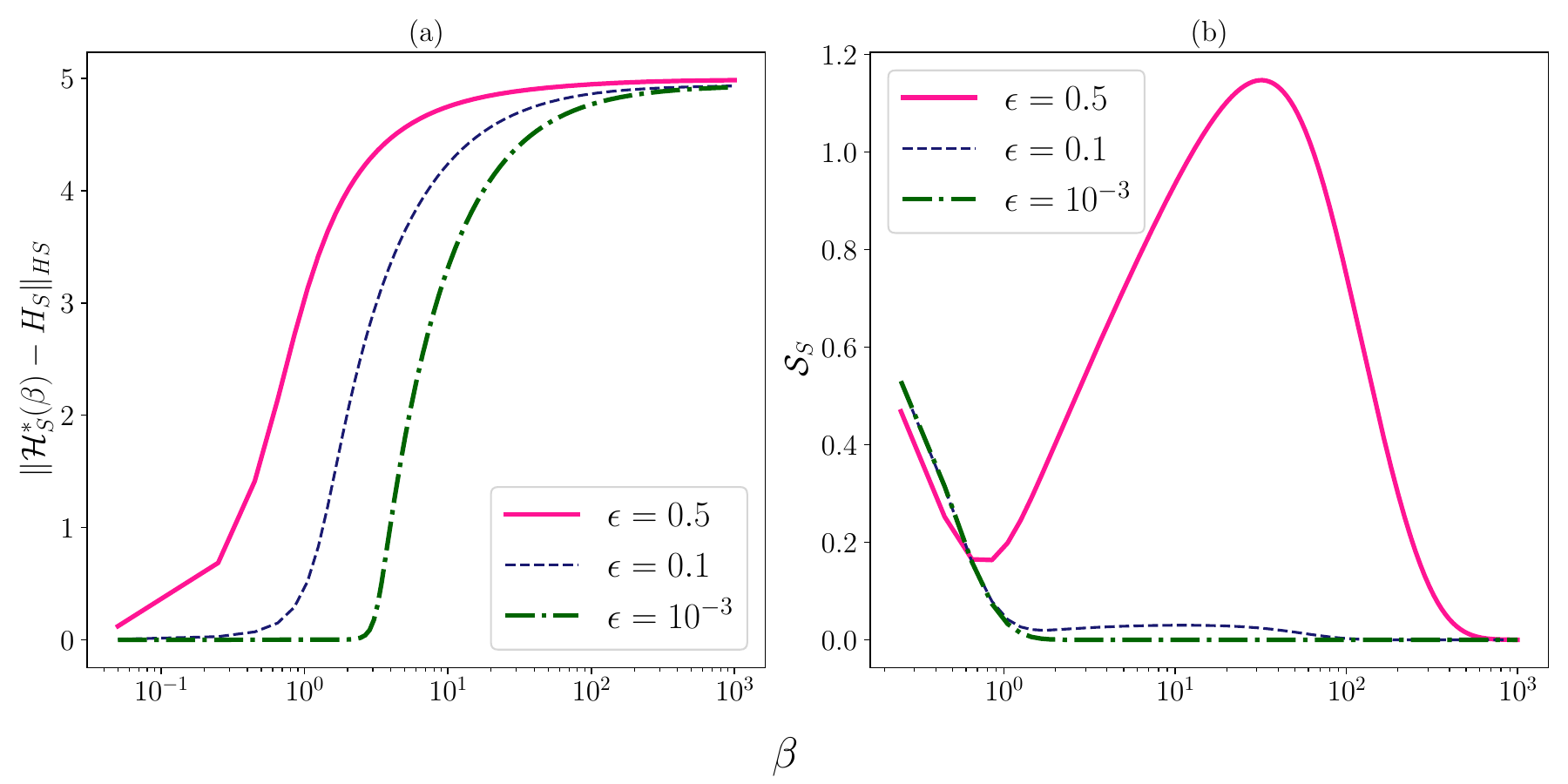}
    \caption{Variation of (a) the Hilbert-Schmidt norm of the matrix $\mathcal{H}_S^*(\beta) - H_S$, where $H_S$ is the central spin system Hamiltonian and $\mathcal{H}_S^*(\beta)$ is the HMF, and (b) thermodynamic entropy $\mathcal{S}_S$, Eq.~\eqref{eq_thermodynamic_entropy_cs} as a function of inverse temperature $\beta$. The parameters are: $N = 80, \omega = \omega_0 = 5$.}
    \label{fig_entropy_hs_norm_hmf_central_spin_model}
\end{figure*}

In Fig.~\ref{fig_entropy_hs_norm_hmf_central_spin_model}(a), we show the impact of interaction strength and bath temperature on the HMF (note that here we have taken a large spin bath $N= 80$) by plotting the Hilbert-Schmidt norm $\left(\|O\|_{HS} = \sqrt{{\rm Tr} [O^\dagger O]}\right)$ of the difference between the system's bare Hamiltonian $H_S$ and HMF $\mathcal{H}_S^*(\beta)$ as a function of inverse temperature $\beta$ and for different system-bath interaction strengths $\epsilon$. It can be observed from the plots that the HMF becomes equal to the system's bare Hamiltonian at high temperatures and in the weak coupling regime. The difference is higher for higher $\epsilon$ and lower temperatures. However, the difference saturates as we keep decreasing the temperature. The thermodynamic entropy of the system is plotted in Fig.~\ref{fig_entropy_hs_norm_hmf_central_spin_model}(b) for different $\epsilon$ as a function of inverse temperature $\beta$. The thermodynamic entropy takes higher values at low temperatures when the system-bath coupling is higher. As we decrease the system-bath coupling, the entropy modulates more slowly for low-temperature values. At very low temperatures, the entropy can be seen to become zero, consistent with the third law of thermodynamics.

\subsection{The canonical Hamiltonian of the central spin model}\label{sec_canonical_Ham}
Following the pseudo-Kraus operator formalism given in \cite{Hayden_2022,breuer_effective_Ham1}, we can calculate the canonical Hamiltonian of the central spin system, which is given by
\begin{align}\label{CanHam}
    H^{\rm can}_S(t)=\Omega(t)\sigma_z,
\end{align}
where
\begin{align}
    \Omega(t)&=-\frac{1}{d}\left( \Lambda_3(t) \frac{\Im\left[y_3(t)\right]}{1+\left|y_3(t)\right|^2}+\Lambda_4 (t) \frac{\Im\left[y_4(t)\right]}{1+\left|y_4(t)\right|^2} \right).
\end{align}
Here, 
\begin{align}
    \Lambda_3&=\frac{1}{2}\left(\zeta(t)-\Gamma(t)-\sqrt{\left[\Gamma(t)+\zeta(t)\right]^2+4\left|\Theta(t)\right|^2}\right),\nonumber \\
    \Lambda_4&=\frac{1}{2}\left(\zeta(t)-\Gamma(t)+\sqrt{\left[\Gamma(t)+\zeta(t)\right]^2+4\left|\Theta(t)\right|^2}\right),
\end{align}
such that
\begin{align} \label{zetatheta}
     \zeta(t)&=\frac{\Dot{\alpha}(t)[1-\eta(t)]+\Dot{\eta}(t)[\alpha(t)-1]}{\alpha(t)-\eta(t)},\nonumber\\
    \Gamma(t)&=\frac{-\Dot{\alpha}(t)\eta(t)+\Dot{\eta}(t)\alpha(t)}{\alpha(t)-\eta(t)},\nonumber\\
    \Theta(t)&=\frac{\Dot{\delta}(t)}{\delta(t)},
\end{align}
and 
\begin{align}
    y_3(t)=\frac{\zeta(t)+\Gamma(t)-\sqrt{\left[\Gamma(t)+\zeta(t)\right]^2+4|\Theta(t)|^2}}{2\Theta^*(t)},\nonumber\\
    y_4(t)=\frac{\zeta(t)+\Gamma(t)+\sqrt{\left[\Gamma(t)+\zeta(t)\right]^2+4|\Theta(t)|^2}}{2\Theta^*(t)}.
\end{align}
$\Im(\cdot)$ denotes the imaginary part of $(\cdot)$. Further, $\Omega(0) = \frac{\omega_0}{2}$. 
\begin{figure}
    \centering
    \includegraphics[width = 1\columnwidth]{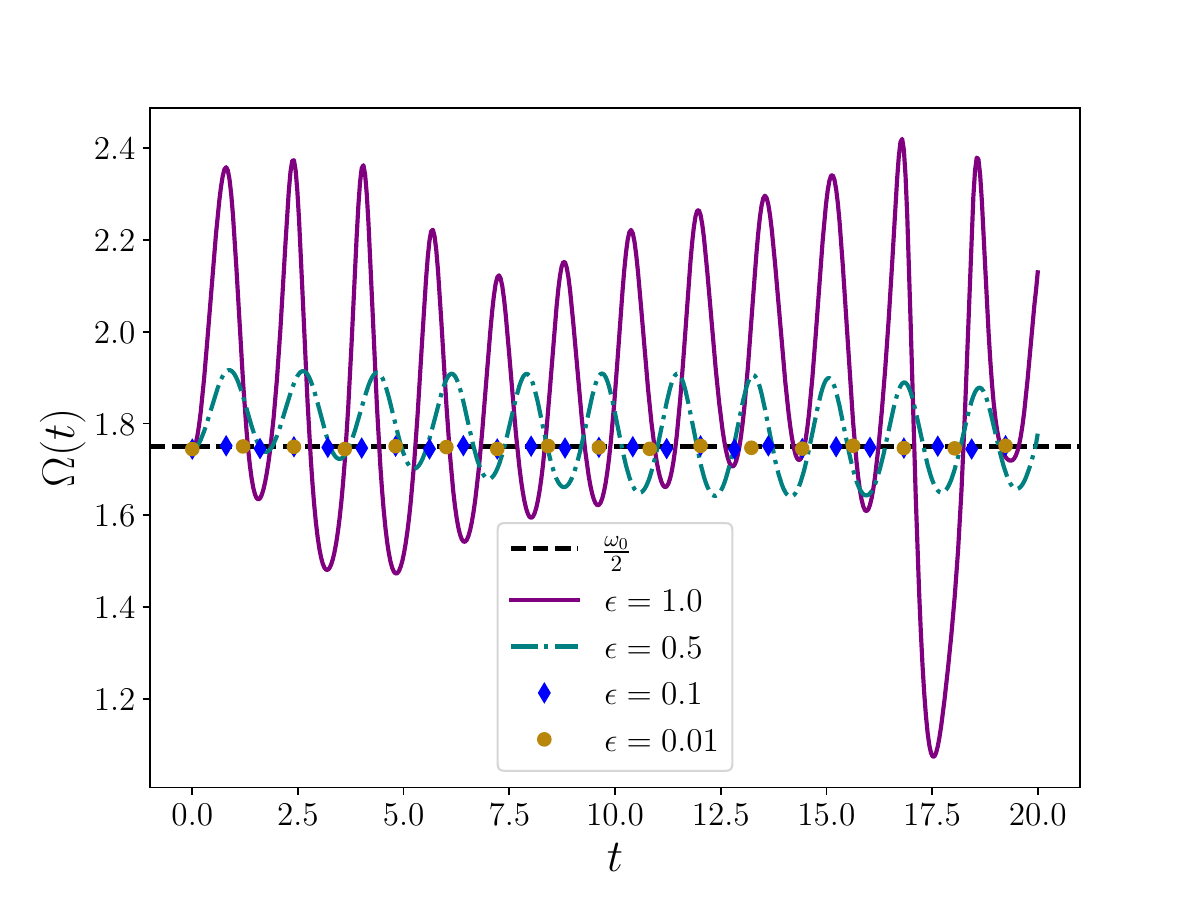}
    \caption{Variation of the factor $\Omega(t)$ present in the canonical Hamiltonian $H^{\rm can}_S(t)$ [Eq. (\ref{CanHam})] with time $t$ for different values of system-bath interaction parameter $\epsilon$. The values of parameters are chosen to be $N = 50, T = 0.1, \omega = 4$, and $ \omega_0 = 3.5$.}
    \label{fig_D_t}
\end{figure}
In Fig. \ref{fig_D_t}, we plot the factor $\Omega(t)$ with time $t$. 
As time progresses, the factor $\Omega(t)$ shows an oscillatory behavior with time around the line of $\frac{\omega_0}{2}$; however, it is asymmetric. The minimal dissipator can be found using Eq.~\eqref{eq:minimal_dissipator}. 

After the CPTP map is applied, partial tracing is done to extract the system's density matrix. This process of tracing out amounts to a loss of the mutual information between the system and the bath and the local information of the bath \cite{landi_RevModPhys.93.035008}. In \cite{Hayden_2022,breuer_effective_Ham1}, it was shown that the canonical Hamiltonian is a unique Hamiltonian when the dissipator in the master equation is minimized with respect to a specific norm in the superoperator space. This canonical Hamiltonian contains the information of both the system and the bath. It effectively drives the system density matrix while in the influence of the bath.

In \cite{nazir_segal_PRXQuantum.4.020307,breuer_napoli_messina_PhysRevB.78.064309}, an effective Hamiltonian is talked about. Through methods like reaction coordinate \cite{nazir_segal_PRXQuantum.4.020307} transformation, the original Hamiltonian is replaced by a mathematically similar effective Hamiltonian such that it has explicit dependence on coupling parameters, allowing for interpretations of strong coupling cases~\cite{Strasberg_2016}. 

\subsubsection{Role of the canonical Hamiltonian in central spin's thermodynamics}\label{sec:role_can_ham}
\begin{figure*}
    \centering
    \includegraphics[width = 1.75\columnwidth]{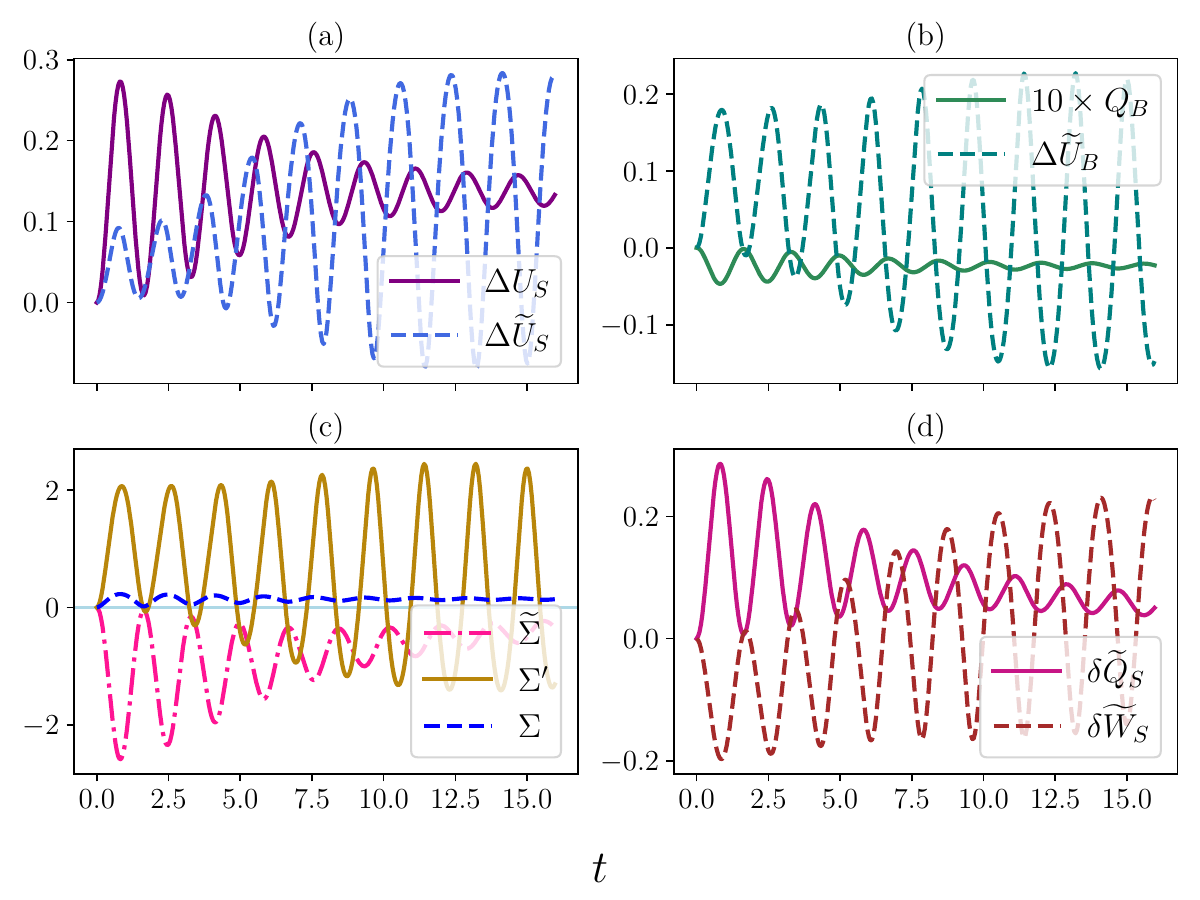}
    \caption{Variation of (a) the change in the internal energy of the system, (b) the change in the internal energy of the bath, (c) the entropy production, and (d) $\delta \widetilde Q_S$ and $\delta \widetilde W_S$ with time. The parameters are chosen to be: $N = 50$, $\omega = 3$, $\omega_0 = 3.5$, $\epsilon = 0.5$, and $T = 0.1$. The system is initially in the thermal state $e^{-\beta H_S}/Z$, where $Z = {\rm Tr}\left(e^{-\beta H_S}\right)$.}
    \label{fig_heat_work_entropy_using_canHam}
\end{figure*}
In the preceding section, we discussed the quantum thermodynamics of the central spin system using the system's bare Hamiltonian $H_S$ and the reduced state of the system at any time $t$. However, the dynamics of the reduced state of the system is obtained from a quantum master equation having a canonical Hamiltonian $H^{\rm can}_S$, as discussed above. Therefore, it becomes natural to discuss the role of the canonical Hamiltonian in calculating the thermodynamical observables~\cite{breuer_effective_Ham1, Colla_Breuer_otto_PhysRevResearch.6.013258}. To this end, we calculate the change in energy of the system by replacing the system's Hamiltonian with the canonical Hamiltonian, that is, 
\begin{align}
    \Delta \widetilde U_S (t) = {\rm Tr} \left[H^{\rm can}_S(t) \rho_S(t)\right] - {\rm Tr}\left[H^{\rm can}_S(0) \rho_S(0)\right].
\end{align}
The change in internal energy of the system can be broken into two parts, which are $\delta \widetilde W_S(t)$ and $\delta \widetilde Q_S(t)$. The expressions for these quantities are given by
\begin{align}\label{eq_can_ham_work}
    \delta \widetilde W_S(t) &= \int_0^t d\tau ~~{\rm Tr} \left[\dot H^{\rm can}_S(\tau)\rho_S(\tau)\right], \nonumber \\
    \delta \widetilde Q_S(t) &= \int_0^t d\tau ~~{\rm Tr}\left[H^{\rm can}_S(\tau)\dot \rho_S(\tau)\right].
\end{align}
Here, the relation $\Delta \widetilde U_S(t) = \delta \widetilde W_S(t) + \delta \widetilde Q_S(t)$ resembles the first law of thermodynamics. 

One may argue that the canonical Hamiltonian contains the effect of the bath even when the bath is partially traced and that it should be used to calculate expectation values of the thermodynamical quantities. By this argument, we observe that the system's Hamiltonian becomes time-dependent even if we take a time-independent initial Hamiltonian of the system. We have already seen in the previous section that the work done on the system can be calculated as the mismatch between the internal energies of the system and the bath. However, we observe here that by using the canonical Hamiltonian, we can get non-zero work solely due to the time dependence of the system's Hamiltonian, Eq. (\ref{eq_can_ham_work}). This analysis was introduced in~\cite{breuer_effective_Ham1}, where the time dependence of $H^{\rm can}_S(t)$ was held responsible for the exchange of energy between the open system and the bath. As a result of this definition of heat exchange, entropy production can be defined by 
\begin{align}\label{Breuers_entropy_production}
    \widetilde \Sigma = \Delta S_S(t) - \beta \delta \widetilde Q_S(t),
\end{align}
where $\Delta S_S(t)$ is given in Eq. (\ref{entropy_prodcution_landi_thermal}). It is interesting to note here that upon comparing the above equation with Eq. (\ref{entropy_prodcution_landi_thermal}), the quantities $Q_B$ and $\delta \widetilde Q_S(t)$ seem to play a similar role, albeit with opposite signs. 

We take a moment here and investigate whether tracing out the system has any effect on the bath, that is, whether the bath's Hamiltonian also changes in a similar way as the system's Hamiltonian changes. This may not happen when the bath is too big to be affected by the operation of tracing out the system. However, in the case of a finite bath, which we have considered here, the bath Hamiltonian may also change. To this end, we need to modify the definition of the change in the internal energy of the bath by replacing $H_B$ in Eq. (\ref{Q_B}) with a time-dependent one, such that
\begin{align}
     \Delta \widetilde U_B (t) = {\rm Tr}\left[\widetilde H_B(t) \rho_B(t)\right] - {\rm Tr}\left[\widetilde H_B(0) \rho_B(0)\right].
\end{align}
But to calculate the time-dependent Hamiltonian $\widetilde H_B(t)$ of the bath, we need to know the map for the evolution of the bath, which is a highly non-trivial task. Nevertheless, we can adopt a different way of calculating the change in the energy of the bath $\Delta \widetilde U_B(t)$. Analogous to Eq.~\eqref{bath_heat_current_bare}, we can compute the bath heat current in the present scenario. Previously, we conditioned the bath Hamiltonian to be time-independent and derived the expression for the change in the internal energy of the bath. In the present scenario, the rate of change in the internal energy of the bath is given by $\Delta \dot{\widetilde{U}}_B  = \frac{d}{dt}\left\langle\widetilde H_B (t)\right\rangle$. Again, using the relation between the expectation of any quantum mechanical operator and the expectation value of the commutator between the operator and the total Hamiltonian of the system, we can rewrite $\Delta \dot{\widetilde{U}}_B$ as
\begin{align}\label{eq_delta-dot-ub}
    \Delta \dot{\widetilde{U}}_B = -i\left\langle[\widetilde H_B(t), V]\right\rangle + \left\langle\frac{\partial \widetilde H_B(t)}{\partial t}\right\rangle.
\end{align}
Invoking the fact that the interaction and the total Hamiltonians are still time-independent, we get \begin{align}
\frac{\partial H^{\rm can}_S(t)}{\partial t} = -\frac{\partial \widetilde H_B(t)}{\partial t}.
\end{align}
Furthermore, along similar lines as above, the rate of change in the expectation value of $\left\langle V(t)\right\rangle$ is given by 
\begin{align}\label{eq_ddt_V}
    \frac{d}{dt} \left\langle V(t)\right\rangle = -i\left\{\left\langle[V, H^{\rm can}_S(t)]\right\rangle + \left\langle[V, \widetilde H_B(t)]\right\rangle\right\},
\end{align}
and the rate of change in the internal energy of the system can be written as 
\begin{align}\label{eq_delta_dot_uS}
    \Delta \dot {\widetilde{U}}_S(t) = \frac{d}{dt}\left\langle H^{\rm can}_S(t)\right\rangle = -i\left\langle[H^{\rm can}_S, V]\right\rangle + \left\langle\frac{\partial H^{\rm can}_S (t)}{\partial t}\right\rangle.
\end{align}
Notice that the first term on the extreme right side of the above equation is equal to $\delta \dot{\widetilde{Q}}_S(t)$, and the second term is equal to the $\delta \dot{\widetilde{W}}_S(t)$. 
Further, by combining the above equations, we can rewrite the expression for the change in the internal energy of the bath $\Delta \widetilde U_B(t)$ in the presence of a time-dependent bath Hamiltonian, which is given by 
\begin{align}\label{eq_first_law_using_canHam}
    \Delta \widetilde U_B(t) = {\rm Tr}\left[V\left\{\rho_{SB}(0) - \rho_{SB}(t)\right\}\right] - \Delta \widetilde U_S(t).
\end{align}
Interestingly, the above equation matches the form of the first law of thermodynamics provided in Eq. (\ref{eq_first_law}). In this case, the mismatch in the change in the internal energies of the system $\Delta \widetilde U_S(t)$ and the bath $\Delta \widetilde U_B(t)$ is balanced by the first term on the right side. This term is the same as in Eq. (\ref{work_done_by_bath}) for the work done by the bath. Therefore, even in the case of a time-dependent system and bath Hamiltonian and for constant interaction and total Hamiltonian, the work done by the bath on the system is given by Eq. (\ref{work_done_by_bath}).
Having defined the above thermodynamic quantities for a time-dependent system and bath Hamiltonian, it would be interesting to construct a quantity similar to entropy production using $ \Delta \widetilde  U_B(t)$, inspired by Eq.~\eqref{entropy_prodcution_landi_thermal}. We call it $\Sigma'$, where
\begin{align}\label{eq_entropy_production_bath_time}
    \Sigma' = \Delta S + \beta \Delta \widetilde U_B(t).
\end{align}

\begin{figure*}
    \centering
    \includegraphics[width = 1.75\columnwidth]{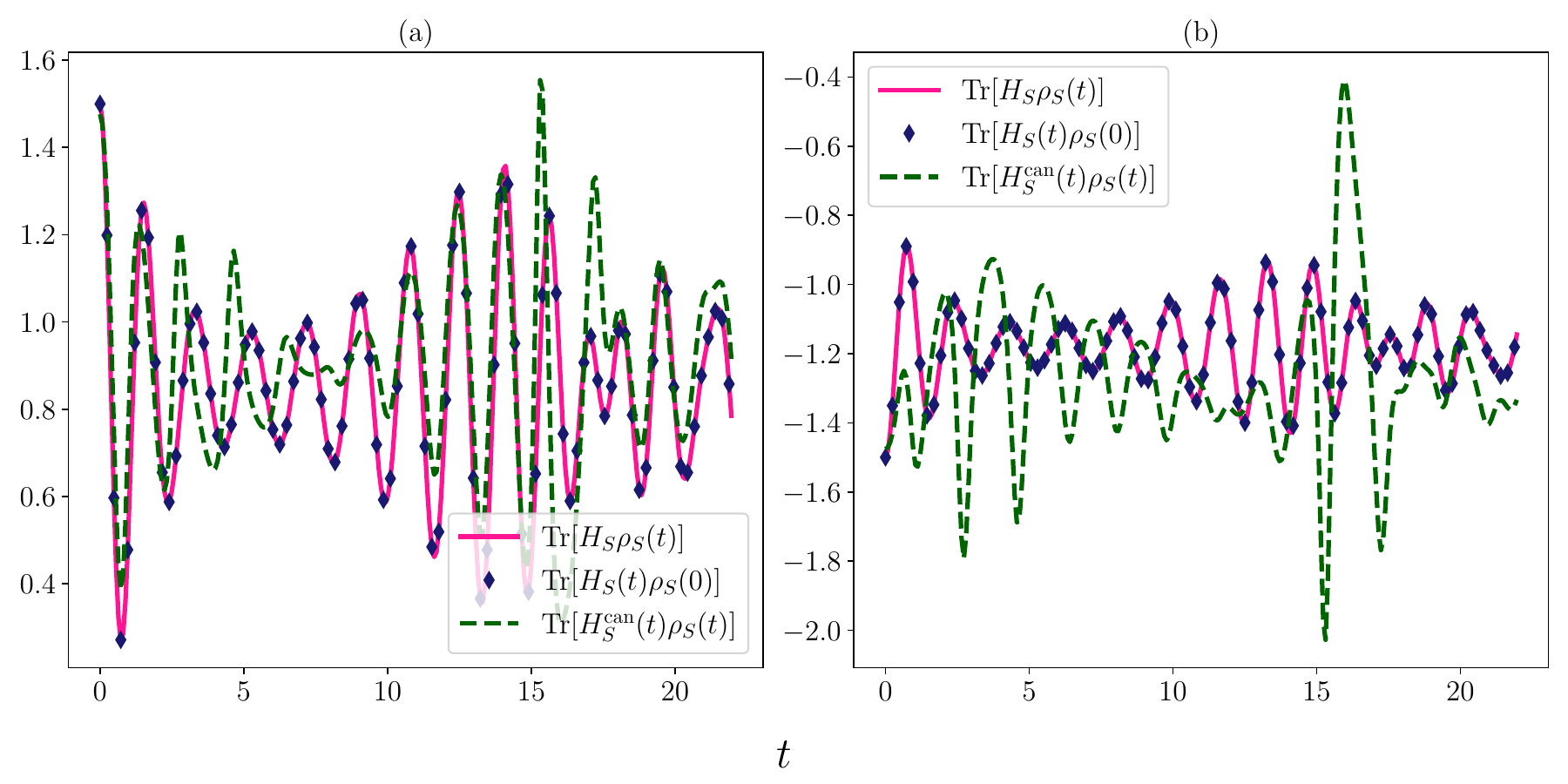}
    \caption{Variation of expectation value $\braket{H_S}$ using Schr\"{o}dinger picture $\braket{H_S} = {\rm Tr}\left[H_S\rho_S(t)\right]$ and Heisenberg picture $\braket{H_S} = {\rm Tr}\left[H_S(t)\rho_S(0)\right]$, along with the analogous expressing using the canonical Hamiltonian ${\rm Tr}\left[H_S^{\rm can}(t)\rho_S(t)\right]$ with time. The parameters are: $\omega = 3.5, \omega_0 = 3.0, \epsilon = 1.0, N = 50$ and $T = 0.10$. The initial state is taken to be (a) the excited state and (b) the ground state.}
    \label{fig:schrodinger_heisenberg_equivalence}
\end{figure*}

The thermodynamic quantities computed in this section using the canonical Hamiltonian are plotted in Fig.~\ref{fig_heat_work_entropy_using_canHam}, along with the corresponding quantities from the previous section to benchmark their variations. From an immediate inspection, we see that the quantities calculated using the canonical Hamiltonian show enigmatic behavior. The entropy production, Fig.~\ref{fig_heat_work_entropy_using_canHam}(c), can be observed to be negative when computed using Eq. (\ref{Breuers_entropy_production}), violating the second law of thermodynamics. This equation of entropy production has two problems. First, it is calculated in the system's subspace, albeit considering the canonical Hamiltonian carries the effect of the bath even after it is traced out. A similar observation of this expression of entropy production being negative was also observed in~\cite{Esposito_2010}. The other issue is that even if we replace the canonical Hamiltonian with the system's Hamiltonian and further take it to be time-independent, the expression for the entropy production $\Sigma$ from Eq. (\ref{entropy_prodcution_landi_thermal}) doesn't match Eq. (\ref{Breuers_entropy_production}) due to the presence of work done by the bath $W$. Furthermore, it is observed from Fig.~\ref{fig_heat_work_entropy_using_canHam}(c) that the entropy production $\Sigma$ in the form of Eq.~(\ref{entropy_prodcution_landi_thermal}) preserves the second law of thermodynamics ($\Sigma \ge 0$) for the same set of parameters. We also plot the quantity $\Sigma'$, Eq. (\ref{eq_entropy_production_bath_time}), where we replace the internal energy change of the bath $Q_B$ in Eq. (\ref{entropy_prodcution_landi_thermal}) by the internal energy change of the bath due to the time-dependent bath Hamiltonian $\Delta \widetilde U_B(t)$. However, this quantity also goes negative at various points in time. We refrain from calling it an expression of entropy production because even if we consider a time-dependent bath Hamiltonian, the entropy production [Eq. (\ref{entropy_production})], for the initial thermal state of the bath, results in Eq. (\ref{entropy_prodcution_landi_thermal}) and not in Eq. (\ref{eq_entropy_production_bath_time}). 

The involvement of the canonical Hamiltonian in calculating the heat exchange and work done further poses an issue with the first law of thermodynamics. In this section, we have seen that there are two possible definitions of the first law of thermodynamics. In the first, $\delta \widetilde W_S(t)$ and $\delta \widetilde Q_S(t)$ from Eq. (\ref{eq_can_ham_work}) can be clubbed together to represent the first law of thermodynamics. Here, non-zero work is obtained because of the canonical Hamiltonian $H^{\rm can}_S(t)$. However, from Figs.~\ref{fig_heat_work_entropy_using_canHam}(a) and (b), it can be observed that at the start of the dynamics, the change in the internal energy of the system $\Delta \widetilde U_S(t)$ is positive, and so is the change in the internal energy of the bath $\Delta \widetilde U_B(t)$. Further, in the same temporal regime, the work done $\delta \widetilde W_S$ is negative, Fig.~\ref{fig_heat_work_entropy_using_canHam}(d), depicting that the system works on the bath, and the heat is transferred to the system as $\delta \widetilde Q_S(t)$ is positive. This observation reveals that on using the canonical Hamiltonian, we get a scenario where the system and the bath both gain energy, and the system works on the bath while heat is transferred to the system. This scenario is ambiguous as it requires some external source to supply energy, which is absent here. The other definition of the first law of thermodynamics comes from Eq. (\ref{eq_first_law_using_canHam}). In this scenario, again, we have the same problem where the system and the bath both gain energy at the same time. However, the work done by the bath, which is the mismatch between the system and the bath energies $W = \Delta \widetilde U_S(t) + \Delta \widetilde U_B(t)$, is initially positive when $\Delta \widetilde U_S(t)$ and $\Delta \widetilde U_B(t)$ are positive. In this case, the work is done by the bath on the system, in contradiction to the work done $\delta\widetilde  W_S$ being negative, as seen above. Therefore, there is an ambiguity in the definition of the work, too. These issues make the formulation of the first law using the canonical Hamiltonian ill-defined.

The above ambiguities resolve when we consider the original time-independent system Hamiltonian $H_S$, Eq.~\eqref{Hname}, for calculating the expectation values of the thermodynamic quantities. In this case, as observed in the previous section, the system gains energy while the bath loses energy, and the mismatch between them is the work done by the bath on the system, which is positive. This scenario is consistent with the first law of thermodynamics. Further, on considering $H_S$ as the system Hamiltonian during the dynamics, we get $\delta \widetilde W_S = 0$, and $\delta \widetilde Q_S(t) = \Delta \widetilde U_S(t) = \Delta U_S$. Therefore, the ambiguity regarding the work done by the bath also resolves and is given by Eq.~(\ref{work_done_by_bath}). Moreover, using the time-independent system and bath Hamiltonians $H_S$ and $H_B$, respectively, the entropy production is also positive, in consonance with Eq.~(\ref{entropy_prodcution_landi_thermal}) that considers irreversibility arising due to tracing out the bath. This scenario is consistent with the second law of thermodynamics. 

The case discussed above can be thought of from a different perspective: when we trace out the bath, the reduced state of the system, which has evolved in time, carries the effect of the bath at any time. After tracing out the bath, the expectation values of the thermodynamic quantities should be calculated using the original system and bath Hamiltonians, which are initially time-independent here. 

To this end, we show in Fig.~\ref{fig:schrodinger_heisenberg_equivalence}, the energy of the reduced system at any time, calculated using the expression $\braket{H_S} = {\rm Tr}\left[H_S\rho_S(t)\right]$ in the Schr\"{o}dinger picture, where the state $\rho_S(t)$ is evolved. At the same time, we plot the expectation value $\braket{H_S} = {\rm Tr}\left[H_S(t)\rho_S(0)\right]$ using the Heisenberg picture. Here, $H_S(t)$ is given by 
\begin{align}
    H_S(t) = {\rm Tr}_B \left[\rho_B(0) \left\{e^{iHt}\left(H_S \otimes \mathbb{I}_B\right) e^{-iHt}\right\}\right],
\end{align}
where $H$ is the total Hamiltonian, as given in Eq.~\eqref{Hname}.
The equivalence of both plots validates the above statement regarding the expectation value of thermodynamic quantities in the two pictures. This can be used to benchmark the expectation value computed using the canonical Hamiltonian, as depicted by the green-dashed curve in the figures, where its non-equivalence to the expectation values from the Schr\"{o}dinger and Heisenberg pictures is evident. The mismatch with the canonical Hamiltonian is much greater when the system is initially in the ground state as compared to when it starts in the excited state. Hence, the canonical Hamiltonian technique does not appear to be suitable for the calculation of quantum thermodynamic quantities.

\section{Ergotropy: application of the central spin model as a quantum battery}\label{sec_ergotropy}
\begin{figure*}
    \centering
    \includegraphics[width = 1.75\columnwidth]{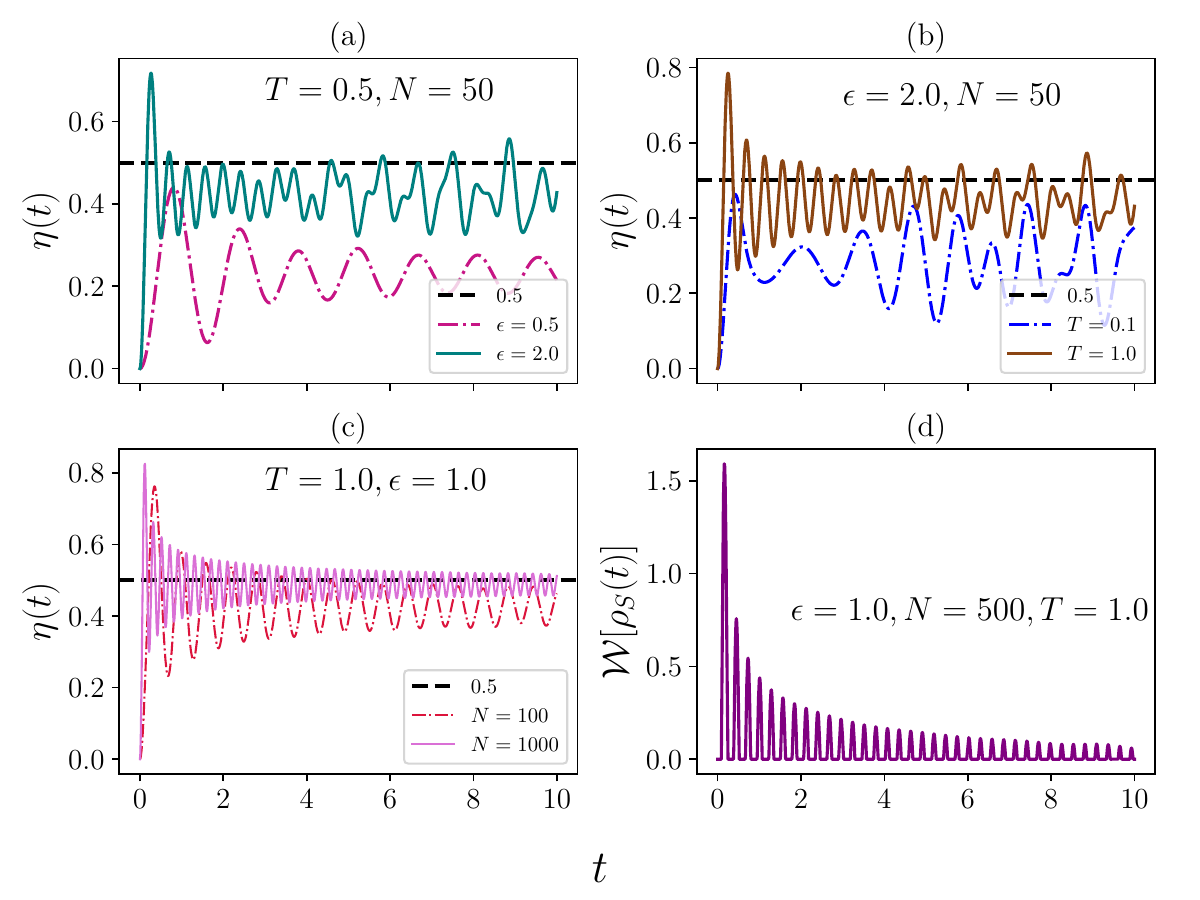}
    \caption{Variation of $\eta(t)$ from Eq. (\ref{eq_etat}) and ergotropy $\mathcal{W}[\rho_S(t)]$ of the central spin system with time $t$ for initial ground state. Here, $\omega_0 = 2.5$, and $\omega = 2.0$. The black-dashed line corresponds to $\eta(t) = 0.5$.} 
    \label{fig_etat}
\end{figure*}%
The maximum amount of work that can be extracted from a quantum system under the influence of some time-dependent cyclic potential is called the ergotropy of the system~\cite{Allahverdyan_2004}
\begin{align}\label{Eq_ergotropy}
    \mathcal{W}[\rho_S(t)]={\rm Tr}\left[\rho_S(t) H_S\right]-{\rm Tr}\left[\rho_S^p H_S\right].
\end{align}
Here, $H_S$ is the system Hamiltonian, and $\rho_S^p$ is a final state that evolved from a state potent of work production $\rho_S(t)$, that is, the passive state. After an external time-dependent cyclic potential has done the maximum work, the state has turned into a passive state, which would be, in general, different from the Gibbs state attained for macroscopic systems. Once the passive state is reached, no more work can be extracted from it through any cyclic process. 
The spectral decompositions of the passive state and the Hamiltonian of the system are given by
\begin{align}
    \rho_S^p=\sum_j r_j \ket{\epsilon_j}\bra{\epsilon_j},\nonumber \\
    H_S=\sum_k \epsilon_k \ket{\epsilon_k}\bra{\epsilon_k},
\end{align}
where the eigenvalues $r_j$ and $\epsilon_j$ have the following ordering:
$r_1\geq r_2 \geq r_3 ... \text{  as  } \epsilon_1 \leq \epsilon_2 \leq \epsilon_3...$~. Thus, the passive state is defined in such a way that the minimum energy state has the maximum population. 
For the central spin system (or, for any single qubit system with $H_S = \frac {\omega_0}2\sigma_z$~\cite{impact_devvrat_SB}), using the Bloch vector form of the system's state at any time $t$, discussed in Eq. (\ref{eq_Bloch_vector_form}), the expression for the ergotropy is given by  
\begin{align}
    \mathcal{W}[\rho(t)] = \frac{\omega_0}{2}\left[z(t) + \sqrt{x(t)^2 + y(t)^2 + z(t)^2}\right].
\end{align}

The ergotropy can be further divided into its incoherent ($\mathcal{W}_i$) and coherent parts ($\mathcal{W}_C$)~\cite{Goold_coherent_ergo}, such that $\mathcal{W} = \mathcal{W}_i + \mathcal{W}_C$. The incoherent ergotropy can be calculated in a similar manner as the ergotropy of the system by replacing the system's state $\rho_S(t)$ in Eq. (\ref{Eq_ergotropy}) with a dephased state $\rho_S^D(t)$. From Eq. (\ref{eq_Bloch_vector_form}), the dephased state is given by $\rho_S^D(t) = \frac{1}{2}\left\{ [1 + z(t)] \ket{0}\bra{0} + [ 1 - z(t)]\ket{1}\bra{1}\right\}$, where $\ket{0}(\ket{1})$ denotes the excited (ground) state of the system. The passive state $\rho_S^{D, p}$ corresponding to this dephased state depends on the sign of $z(t)$. For the single qubit systems with Hamiltonian $H_S = \frac{\omega_0}{2}\sigma_z$ and with positive $z(t)$, the incoherent ergotropy boils down to $\mathcal{W}_i[\rho_S(t)] = \mathcal{W}[\rho_S^D(t)] = \omega_0z(t)$, which for negative $z(t)$ becomes zero. Thus, the coherent ergotropy $\mathcal{W}_C[\rho_S(t)]$ is equal to the ergotropy $\mathcal{W}[\rho_S(t)]$ for negative $z(t)$ and $\mathcal{W}_C[\rho_S(t)] = \frac{\omega_0}{2}\left[\sqrt{x(t)^2 + y(t)^2 + z(t)^2} - z(t)\right]$ for $z(t)$ being positive.  

Ergotropy serves as an important quantifier for the charging and discharging behavior of the system. The system is said to be charging (discharging) when its ergotropy increases (decreases). To this end, the central spin can be visualized as a quantum battery, with the spin bath acting as a charger. A similar analysis in the case of open quantum systems with the system being surrounded by a bosonic bath has been done in~\cite{impact_devvrat_SB, bhanja_devvrat_SB}, where the system was initially taken to be in a positive ergotropy state and the bath acted as a re-charger (due to non-Markovian evolution). However, in the case of the spin bath considered here, we seek whether the bath can act as a charger, that is, whether the bath itself can lead to a state giving some positive ergotropy, even if we take an initial state with zero ergotropy.  To check this, let us consider the central spin system to be initially in a ground state $\rho_S (0) = \ket{1}\bra{1}$, which is a passive state, from Eq. (\ref{Eq_ergotropy}). For this initial state, the factors $x(t)$ and $y(t)$ from Eq. (\ref{eq_Bloch_vector_form}) remain zero at any time $t$. The ergotropy $\mathcal{W}[\rho_S(t)]$ in this case is given by 
\begin{align}
    \mathcal{W}[\rho_S(t)] = \frac{\omega_0}{2}\left[z(t) + \left|z(t)\right|\right]~~~\text{for}~~\rho_S(0) = \ket{1}\bra{1}.
\end{align}
In the case of the ground state, the value of $z(0) = -1$, and therefore, $\mathcal{W}[\rho_S(0)] = 0$. However, the factor $z(t)$ can become positive, thereby making ergotropy positive as time progresses. The condition on $z(t)>0$ leads to $\eta(t)> \frac{1}{2}$, with the help of Eq. (\ref{xt_yt_zt}). 

Figure~\ref{fig_etat} shows the variation of the $\eta(t)$ with respect to time. It is observed that  $\eta(t)$ crosses the $1/2$ line when the interaction strength $\epsilon$ is increased. Further, it is interesting to note that upon increasing the temperature of the bath, $\eta(t)$ is greater than $1/2$ at multiple times. This suggests that the tendency of the bath to act as a charger increases as the temperature increases. The number of spins in the bath $N$ also favors the role of the bath as a charger, as in this case, too, the number of times $\eta(t)$ is greater than $1/2$ increases with an increase in $N$, as can be seen from Fig.~\ref{fig_etat}(c). Figure~\ref{fig_etat}(d) is an example where the initial state of the system has zero ergotropy, yet the ergotropy revives multiple times throughout the evolution due to the bath acting as a charger.

\section{Conclusion}\label{sec:conclusion}
This work is devoted to understanding the quantum thermodynamics of strongly coupled non-Markovian quantum systems. In this context, correct definitions of the fundamental quantum thermodynamic quantities, such as heat, work, entropy production, and ergotropy, as well as the laws of quantum thermodynamics, need to be carefully examined. We addressed this issue by taking up a non-trivial model of a central spin immersed in a spin bath and derived its exact reduced dynamics. This was then utilized to calculate the above thermodynamic quantities using the original system and bath Hamiltonian, where the work done by the bath is the deficit between the system and bath internal energies. This led to appropriate definitions of various thermodynamic quantities and provided consistency with the laws of thermodynamics. Thus, for example, a positive entropy production, verifying the second law of quantum thermodynamics, was observed. However, the entropy production rate was sometimes negative, which is typical behavior of non-Markovian quantum systems. 
We determined the thermodynamic equilibrium state of the system under strong system-bath coupling using the Hamiltonian of Mean Force (HMF) framework. In the weak coupling and high-temperature regime, the HMF of the system was seen to converge to the bare system Hamiltonian. Additionally, we evaluated the thermodynamic entropy of the system based on the HMF; it was found to be consistent with the third law of thermodynamics. Furthermore, we analyzed the influence of both system-bath coupling strength and temperature on the thermodynamic entropy.
Furthermore, the above thermodynamic quantities were also calculated using the recently developed canonical Hamiltonian approach, which led to inconsistencies in the definition of the first and second laws of thermodynamics. 
An interesting result, relevant to the charging-discharging behavior of the central spin, was obtained using ergotropy. In the present case of the spin bath, we arrived at a situation where the bath itself could act as a charger, such that the system (quantum battery) gained positive ergotropy even if initially the ergotropy was zero. This is in contrast to what was observed earlier for a harmonic oscillator bath acting as a charger and, hence, puts the role of the central spin model as a quantum battery in perspective.
It is hoped that this work will provide a stepping stone towards understanding the quantum thermodynamics of strongly coupled non-Markovian systems.

\section*{Acknowledgements}
D. T. thanks Prof. G. T. Landi and Prof. S. Deffner for the useful discussion during the APS March Meeting 2024. 
\begin{appendix}
\section{The spectral decomposition of the total Hamiltonian}
The total Hamiltonian for the central spin model is given in Eq.~(\ref{Hname}). The spectral decomposition of the total Hamiltonian is given by 
\begin{align}
    H &= \sum_{n=1}^N \lambda_{\pm}(n) \ket{\psi_{\pm}(n)}\bra{\psi_{\pm}(n)} + \left(\frac{\omega + \omega_0}{2}\right)\ket{0}_S\ket{0}_B\bra{0}_S\bra{0}_B \nonumber\\
    &- \left(\frac{\omega + \omega_0}{2}\right)\ket{1}_S\ket{N+1}_B\bra{1}_S\bra{N+1}_B,
\end{align}
where 
\begin{align}
    \ket{\psi_\pm(n)} = \frac{\chi_\pm(n) \ket{0}_S \ket{n}_B + \ket{1}_S \ket{n-1}_B}{\sqrt{1 + \chi_\pm (n)^2}},
\end{align}
and 
\begin{align}
    \lambda_\pm (n) = \frac{\left\{N - (2n - 1)\right\}\omega \pm \sqrt{b^2 + 4a_n^2}}{2N}. 
\end{align}
$\ket{i}$ is the standard computational basis in the above equations. The factor $\chi_\pm(n)$ is given by
\begin{align}
    \chi_{\pm}(n)&=\frac{-b\pm \sqrt{b^2+4a_n^2}}{2a_n},
\end{align}
where $b=\omega-N\omega_0$, and 
\begin{align}
    a_n&=\epsilon\sqrt{N}\sqrt{\frac{N}{2}\biggl(\frac{N}{2}+1\biggr)-\biggl(-\frac{N}{2}+n-1\biggr)\biggl(-\frac{N}{2}+n\biggr)}.
\end{align}
Here, $N$ is the number of bath spins, and $n$ is the eigenvalue(eigenvector) index in the range $1\leq n \leq N$. 
\end{appendix}

\bibliography{reference}
\bibliographystyle{apsrev}

\end{document}